\documentclass[12pt]{article}

\usepackage{newtxtext,newtxmath}

\usepackage{graphicx}
\usepackage{booktabs}
\usepackage[letterpaper,margin=1in]{geometry}

\renewenvironment{abstract}
	{\quotation}
	{\endquotation}

\date{}

\makeatletter
\renewcommand{\fnum@figure}{\textbf{Figure \thefigure}}
\renewcommand{\fnum@table}{\textbf{Table \thetable}}
\makeatother

\usepackage{scicite}

\usepackage{url}

\def\scititle{
Engineering quantum optical responses of microtubules through tryptophan-network simulations and ultraviolet spectroscopy 
}

\title{\bfseries \boldmath \scititle}

\author{
Lea~Gassab$^{1\ast}$,
Travis~J.~A.~Craddock$^{2\ast}$\and
\small$^{1}$Departments of Biology, Chemistry and Physics \& Astronomy,\\
\small Waterloo Institute for Nanotechnology, University of Waterloo, Waterloo, ON, Canada.\and
\small$^{2}$Departments of Biology and Physics \& Astronomy,\\
\small Waterloo Institute for Nanotechnology, University of Waterloo, Waterloo, ON, Canada.\and
\small$^\ast$Corresponding author. Email: lgassab@uwaterloo.ca (L.G.); travis.craddock@uwaterloo.ca (T.J.A.C.)
}

\begin{document}
\maketitle

\begin{abstract} \bfseries \boldmath

Microtubules host dense ultraviolet-absorbing aromatic networks, suggesting an opportunity to engineer their optical response for biotechnology. Here we assess the feasibility of tuning microtubule fluorescence by combining an excitonic radiative-coupling model with molecular-dynamics-derived microtubule-like assemblies and steady-state absorbance and fluorescence measurements in microplate geometries. Simulations quantify how positional and orientational fluctuations reshape radiative rates and quantum yield, and predict how perturbing the tryptophan network by removing a specific site, adding an extra tryptophan at candidate binding pockets, or using mixed modification fractions can modulate emission. Experiments on porcine tubulin dimers and taxol-stabilized microtubules support these trends: polymerization enhances microtubule quantum yield at 280 nm and yields bounded changes at 295 nm due to scattering, while added L-tryptophan reproducibly quenches microtubules at both wavelengths. Together, theory and experiment provide evidence for chemically addressable tuning of microtubule quantum yield and motivate design rules for engineered microtubule photonics.

\end{abstract}

\noindent\textbf{Short title:} Engineered quantum optics in microtubules\\
\noindent\textbf{Teaser:} Chemical cofactors tune microtubule fluorescence, enabling engineered quantum optical responses for biotechnology.

\section*{Introduction}

Microtubules are essential structural and dynamical elements of the cytoskeleton that support intracellular transport,
coordinate cell division, and participate in cell signaling~\cite{desai1997microtubule,akhmanova2015control}. In neurons, they
play central roles in maintaining cell architecture, enabling long range axonal transport, and establishing and
remodeling neuronal polarity~\cite{Kelliher2019,Guedes-Dias2019,Iwanski2023,vanBeuningen2016}. Microtubules self-assemble from
$\alpha\beta$ tubulin protein heterodimers~\cite{nogales1998erratum} which exhibit complex polymerization dynamics governed by
nucleotide binding and hydrolysis, and interaction with a variety of structural and motor proteins~\cite{desai1997microtubule}. For these reasons, a central aim of bio-nanotechnology is to harness and modulate microtubule dynamics through novel chemical and optical strategies for cellular control and measurement \cite{akter2024localized, steinbrink2024biological, ishii2022kinesin, bachand2015microtubule, malcos2011engineering}.

Recently, the ultraviolet (UV) optical properties of aromatic amino acids in microtubules have been explored as natural photonic systems that may be used for biotechnological applications. The highly ordered protein scaffold provided by microtubules creates a regular and repeating arrangement of naturally occurring aromatic amino acids, providing unique optical properties and an opportunity for engineered optical functionality \cite{craddock2014feasibility, celardo2019existence}. Theoretically, aromatic lattices in microtubules can support coherent excitation dynamics and related photonic
phenomena~\cite{craddock2014feasibility, kurian2017oxidative}. Recent experiments report photoexcitation diffusion on nanometer scales in microtubules
\cite{kalra2023electronic} strengthening interest in excitonic frameworks in cytoskeletal assemblies. From an engineering
perspective, these results raise a concrete feasibility question: can microtubule UV fluorescence be predictably tuned
by manipulating the effective chromophore network?

Beyond energy transport, collective radiative phenomena such as optical superradiance have been proposed as one possible
mechanism for cooperative light-matter interactions in microtubule tryptophan networks
\cite{celardo2019existence,babcock2024ultraviolet,patwa2024quantum}. In a superradiant regime, delocalized excitonic
eigenstates can exhibit enhanced radiative decay rates relative to isolated emitters, which can in turn modify ensemble
fluorescence quantum yield. Existing modeling has shown that these effects can be highly sensitive to disorder,
geometry, and network connectivity, and experimental steady state measurements have been used to motivate and constrain
such scenarios \cite{babcock2024ultraviolet,patwa2024quantum}. Regardless of whether such collective effects play a
physiological role, they provide a testable design hypothesis for microtubule-based photonics: cooperative radiative
couplings should create network-level signatures in quantum yield that can be modulated by targeted perturbations.

However, two gaps remain important for quantitative interpretation and for assessing feasibility of engineering optical
responses. First, most theoretical studies adopt idealized static microtubule structures, while real tubulin assemblies
exhibit structural variability. Second, steady state microplate spectroscopy introduces practical systematics, most
notably wavelength dependent scattering in microtubule absorbance, that complicate absolute quantum yield estimation,
particularly at tryptophan selective excitation near 295~nm.

Here we address these gaps by combining theory and experiment in a way that explicitly targets robust, perturbation-based
trends relevant to engineering. On the theory side, we use a non Hermitian excitonic Hamiltonian to model radiative
couplings and compute thermal quantum yields for microtubule like assemblies built from molecular dynamics snapshots.
This enables a direct assessment of how realistic fluctuations in chromophore positions and orientations suppress or
preserve cooperative radiative enhancement and how the yield scales with microtubule segment length under structural and
energetic disorder. We then evaluate controlled modifications of the tryptophan network, including selective tryptophan
removal, addition of an extra tryptophan at a candidate binding pocket, and mixed modification fractions, to identify
experimental signatures and candidate levers for chemical tuning.

On the experimental side, we perform steady state UV absorbance and fluorescence measurements on porcine tubulin dimers
and taxol stabilized microtubules using UV transparent microplates and a multimode plate reader. We quantify fluorescence
quantum yields under 280 and 295~nm excitation in baseline conditions and under defined Trp to tubulin doping ratios.
Because microtubule absorbance includes a substantial scattering background in the microplate geometry, we treat
microtubule quantum yields, especially at 295~nm, as bounded estimates rather than a single unique value. We therefore
emphasize two complementary comparisons. The first is polymerization induced changes from tubulin dimers to microtubules,
reported using both uncorrected and scattering corrected absorbance branches. The second is tryptophan dependent changes
within the microtubule class itself, which are less sensitive to the absolute absorbance baseline because they compare MT
conditions measured in the same geometry with similar scattering structure. At 295~nm, where tryptophan is more
selectively excited and direct comparison with a Trp only excitonic model is most meaningful, these within MT comparisons
provide the clearest experimental constraints on cooperative optical behavior and on the impact of chemically perturbing
the effective tryptophan network.

First, we present the Results in three steps: (i) theory predictions for microtubule length scaling under realistic
structural variability and static energetic disorder, constraining cooperative enhancement and quantum yield; (ii)
simulation-guided perturbations of the tryptophan network (tryptophan removal and added tryptophan) as candidate design
controls; and (iii) steady-state ultraviolet spectroscopy of tubulin dimers and taxol-stabilized microtubules under 280
and 295~nm excitation, highlighting which comparisons remain robust when absorbance is influenced by scattering. We then
synthesize these findings in the Discussion in the context of feasibility and design rules for engineered microtubule
photonics, and provide full model construction, excitonic/MD methods, perturbation protocols, and spectroscopy/analysis
details in Materials and Methods.

\section*{Results}
\label{sec:results}

\subsection*{Structural variability suppresses superradiance}

\begin{figure} 
	\centering
	\includegraphics[width=0.9\textwidth]{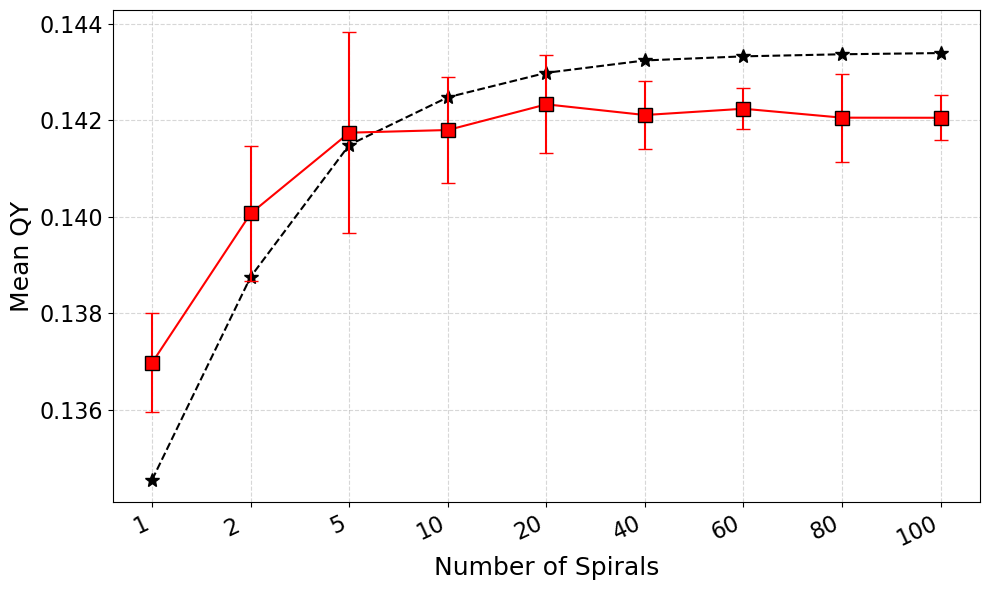}
	\caption{\textbf{Quantum yield increases with microtubule length in molecular-dynamics--derived ensembles.}
	Mean fluorescence quantum yield (QY) versus number of spirals (1--100; 13 tubulin dimers per spiral) for assemblies
	constructed from randomly selected tubulin conformations. Points show the mean of 20 independent realizations; error
	bars show the standard deviation. Black stars indicate the QY for an ordered microtubule constructed from the static
	1JFF structure as reported in~\cite{patwa2024quantum}. Welch two-sided $t$ tests comparing 1 spiral versus 100 spirals
	yield $p=7.10\times10^{-5}$ for the randomized ensemble.}
	\label{fig:rando}
\end{figure}

We built microtubule-like assemblies of varying length by randomly sampling conformations from a 3100-frame molecular dynamics trajectory and assembling segments from 1 to 100 spirals (13 tubulin dimers per spiral). As assembly length increased, the thermal fluorescence quantum yield increased and approached a plateau near 0.142 in the molecular-dynamics–derived ensemble (Fig.~\ref{fig:rando}). This plateau is slightly lower than the ordered reference value reported for the static 1JFF-based structure, and the net change from short to long assemblies is reduced in the randomized ensemble relative to the ordered case (Fig.~\ref{fig:rando}).

\begin{figure} 
	\centering
	\begin{minipage}{0.49\textwidth}
		\centering
		\includegraphics[width=\textwidth]{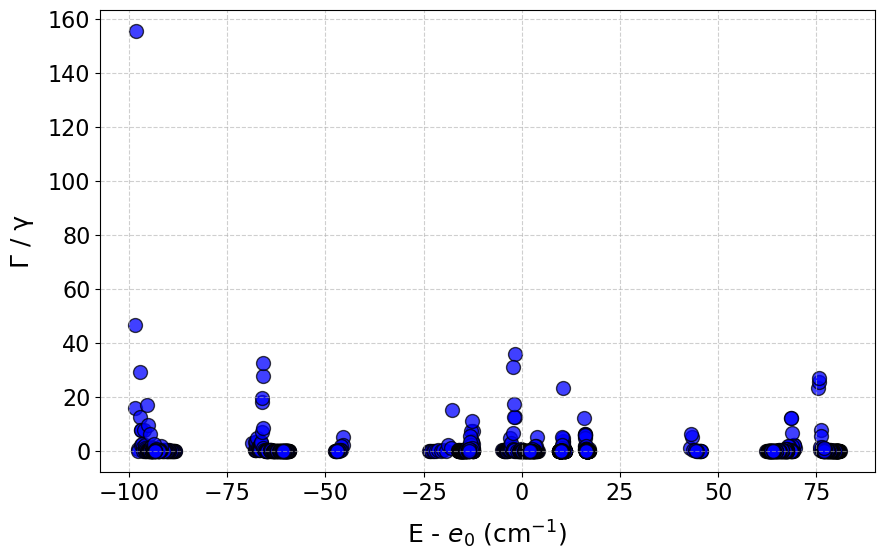}
	\end{minipage}\hfill
	\begin{minipage}{0.49\textwidth}
		\centering
		\includegraphics[width=\textwidth]{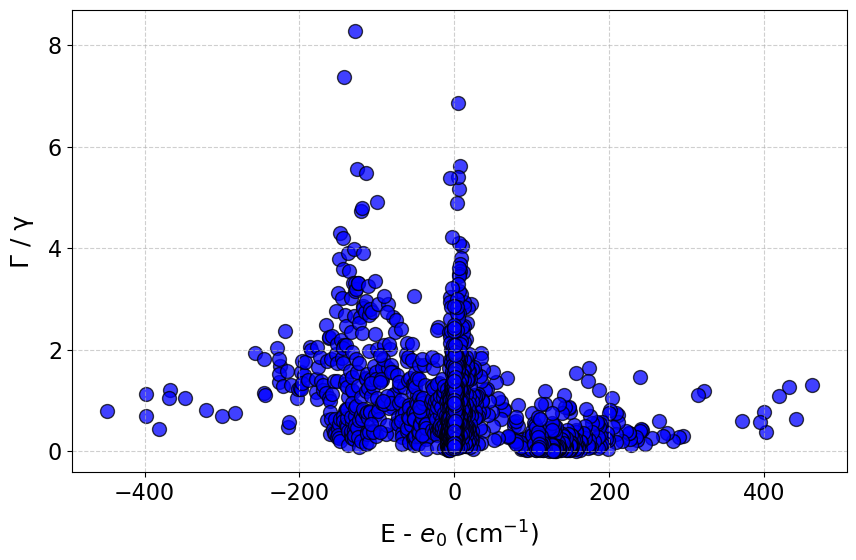}
	\end{minipage}
	\caption{\textbf{Structural variability suppresses maximal superradiant enhancement.}
	Eigenvalue spectra showing radiative-rate enhancement versus energy for microtubule segments in two structural
	regimes. Left, idealized ordered reference constructed from the static 1JFF tubulin structure ($\mathrm{QY}=0.1425$).
	Right, 10-spiral assemblies constructed from randomly selected tubulin conformations from molecular dynamics
	simulations ($\mathrm{QY}=0.1417$).}
	\label{fig:graph_ordered_vs_md}
\end{figure}

We observed a much stronger effect on radiative-rate enhancement than on quantum yield. When we included structural variability, the maximal superradiant enhancement dropped substantially relative to the ordered reference, and the distribution of decay rates across excitation energies shifted (Fig.~\ref{fig:graph_ordered_vs_md}). Despite the reduced maximal enhancement, the molecular-dynamics–derived ensembles maintained quantum yields above 0.14 over the lengths examined.

\subsection*{Diagonal disorder preserves length scaling}

\begin{figure} 
	\centering
	\includegraphics[width=0.9\textwidth]{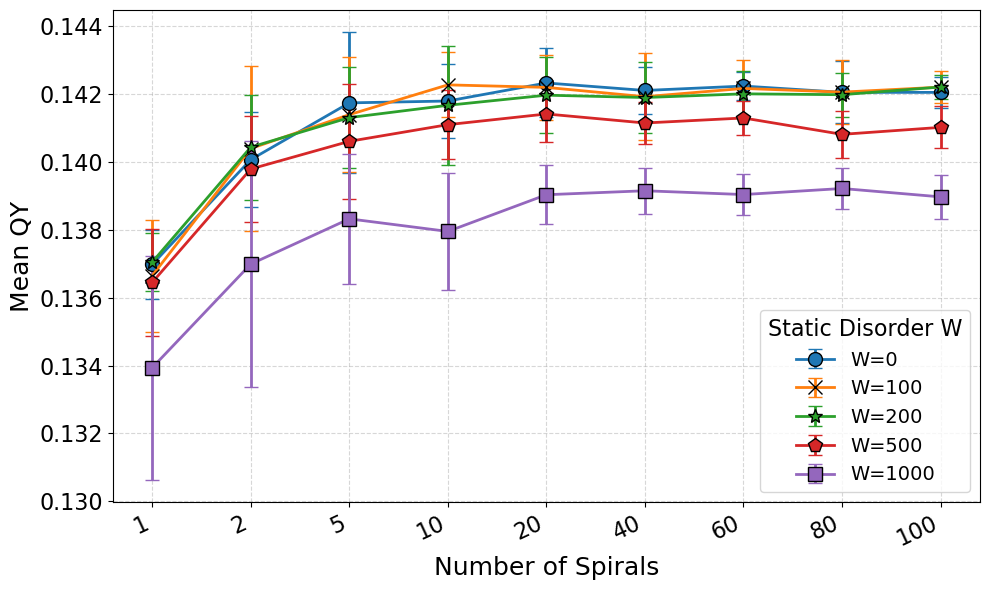}
	\caption{\textbf{Length-dependent quantum yield persists under static energetic disorder.}
	Mean QY versus number of spirals (1--100) for different diagonal disorder strengths $W$ (cm$^{-1}$), where site-energy
	offsets are sampled uniformly from $\left[-W/2,\,W/2\right]$. Points show the mean of 20 independent realizations;
	error bars show the standard deviation. Welch two-sided $t$ tests comparing 1 spiral versus 100 spirals yield
	$p=7.10\times10^{-5}$ ($W=0$), $1.25\times10^{-6}$ ($W=100$), $5.95\times10^{-6}$ ($W=200$), $5.08\times10^{-5}$
	($W=500$), and $1.84\times10^{-3}$ ($W=1000$).}
	\label{fig:graph_ran_W}
\end{figure}

To test robustness to static energetic disorder, we added random site-energy offsets to the Hamiltonian with disorder strength $W$ (cm$^{-1}$); see the Materials and Methods for details. The fluorescence quantum yield remained nearly unchanged up to $W=200$~cm$^{-1}$, decreased at $W=500$~cm$^{-1}$ while remaining above 0.14, and fell below 0.14 at $W=1000$~cm$^{-1}$ (Fig.~\ref{fig:graph_ran_W}). The impact of static disorder in randomly constructed microtubule networks is consistent with the corresponding 1JFF-based results reported previously~\cite{patwa2024quantum}, indicating that length-dependent quantum-yield enhancement persists when random tubulin configurations and diagonal disorder are applied simultaneously. Across all disorder levels examined, Welch two-sided tests comparing 1 spiral versus 100 spirals remained significant, with the largest $p$-value $p = 1.84 \times 10^{-3}$ at $W = 1000$~cm$^{-1}$, confirming that length-dependent yield trends persist even under strong diagonal disorder (Fig.~\ref{fig:graph_ran_W}). While $W=200$~cm$^{-1}$ has often been taken as a representative room-temperature disorder scale based on $k_{B}T$~\cite{patwa2024quantum}, recent measurements in light-harvesting complexes such as photosystem~I report thermal site-energy fluctuations approaching $100$~meV (approximately $800$~cm$^{-1}$)~\cite{reiter2023thermal}. Consistent with this larger fluctuation scale, the quantum yield at $W=1000$~cm$^{-1}$ drops below 0.14, indicating that sufficiently strong energetic disorder can hinder the robustness of quantum-yield enhancement.

\subsection*{Tryptophan perturbations tune quantum yield}

\begin{figure} 
	\centering
	\includegraphics[width=0.9\textwidth]{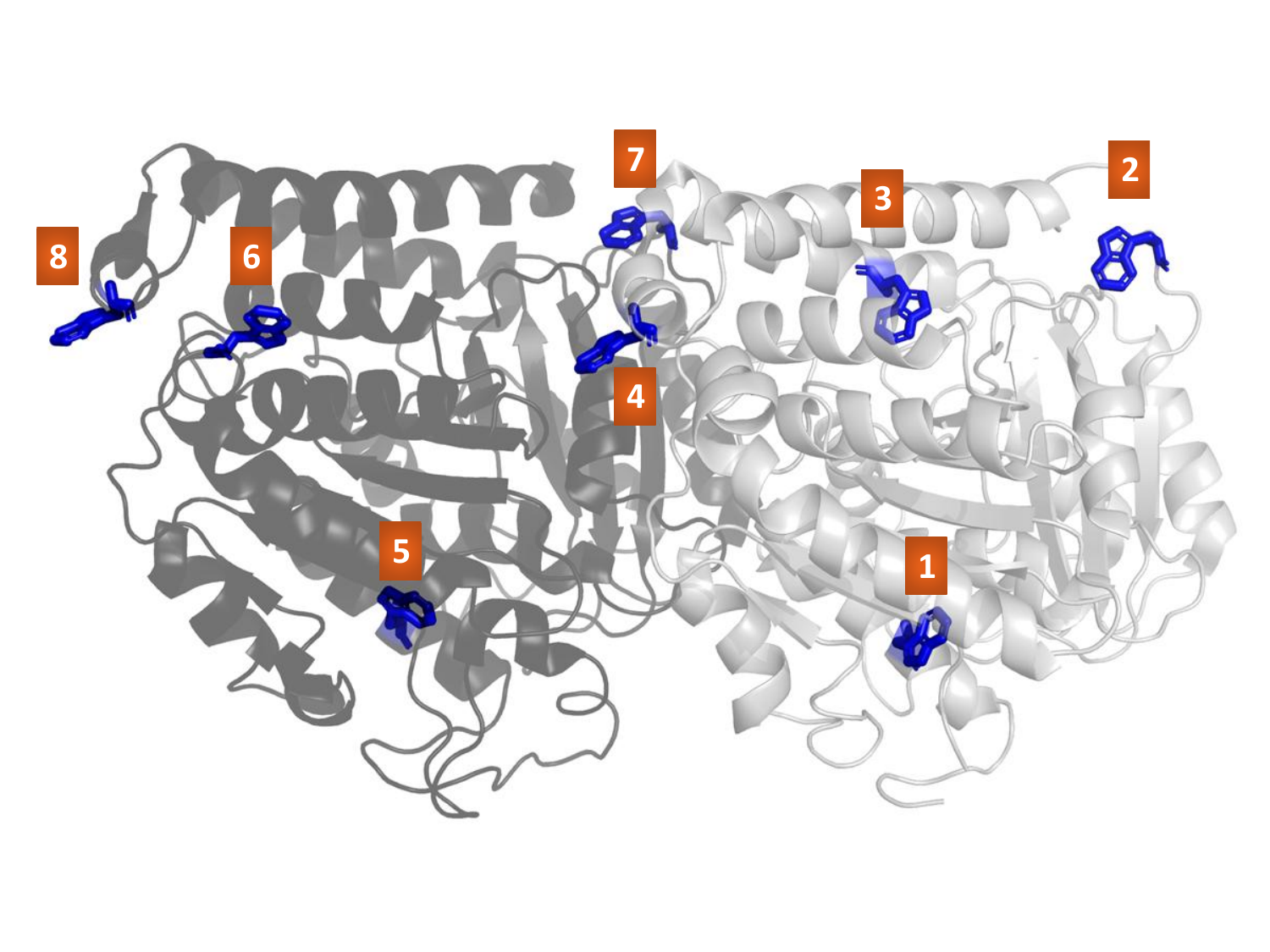}
	\caption{\textbf{Tubulin dimer tryptophan network used for residue mapping.}
	Structure of the $\alpha\beta$-tubulin dimer (PDB 1JFF) rendered in PyMOL~\cite{delano2002pymol}, highlighting the
	eight tryptophan residues. The $\alpha$-tubulin chain is shown in light gray and the $\beta$-tubulin chain in dark
	gray; tryptophans are shown in blue. Red labels indicate the residues used throughout the manuscript: Trp1
	($\alpha$21), Trp2 ($\alpha$346), Trp3 ($\alpha$388), Trp4 ($\alpha$407), Trp5 ($\beta$21), Trp6 ($\beta$103), Trp7
	($\beta$346), and Trp8 ($\beta$407).}
	\label{fig:tub}
\end{figure}

\begin{figure} 
	\centering
	\includegraphics[width=0.9\textwidth]{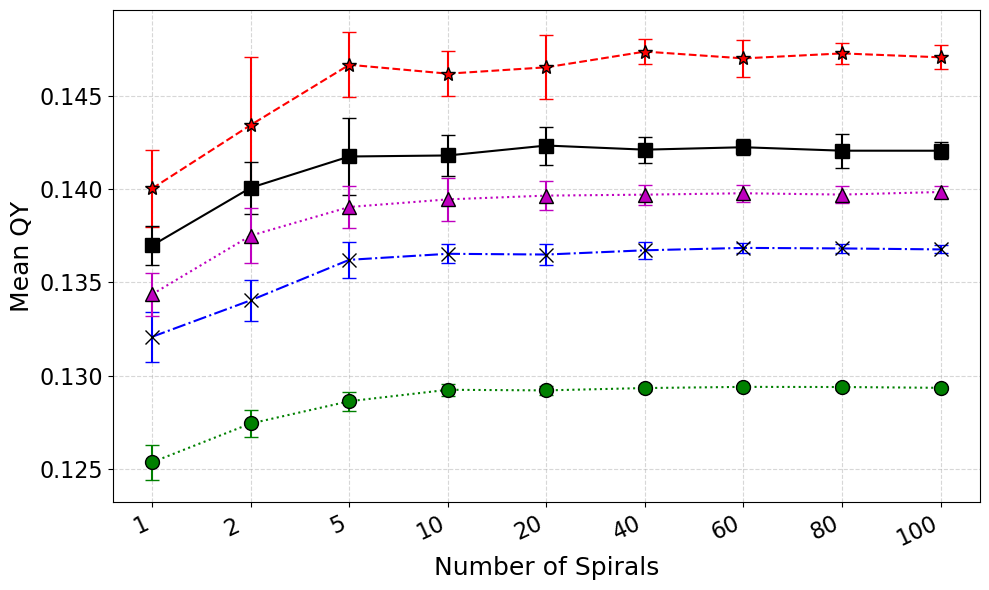}
	\caption{\textbf{Tryptophan-network perturbations tune the predicted quantum yield.}
	Mean QY versus number of spirals (1--100) for assemblies constructed from randomly selected molecular-dynamics
	conformations. Unmodified tubulin (black squares), Trp~3 removed (red stars), added tryptophan at a candidate binding
	pocket (green circles), and mixed assemblies containing 25\% (purple triangles) or 50\% (blue crosses) modified
	dimers. Points show the mean of 20 independent realizations; error bars show the standard deviation.}
	\label{fig:trp_rando}
\end{figure}

We evaluated two perturbation classes in the simulations that map onto experimentally accessible manipulations. First, removal of Trp~3, the tryptophan labeled site~3 on tubulin (see Fig.~\ref{fig:tub}), increased the predicted quantum yield relative to unmodified assemblies, with a representative increase of $\sim$3.5\% (Fig.~\ref{fig:trp_rando}). Second, addition of a single tryptophan near a candidate binding pocket decreased the predicted quantum yield, and this effect strengthened as the fraction of modified dimers increased in mixed assemblies (Fig.~\ref{fig:trp_rando}). The perturbation-averaged predictions for the added-tryptophan case are used for direct comparison to experiment, giving a quantum-yield change of $-1.6\%$ at a 25\% modified fraction and $-8.9\%$ for full modification (Table~\ref{tab:delta_mt}).

\subsection*{Spectroscopy separates robust comparisons}
\label{sec:results_steadystate}

We measured ultraviolet absorbance and steady-state fluorescence for tubulin dimers (TuD) and taxol-stabilized microtubules (MT) under 280 and 295~nm excitation in three conditions: no added tryptophan (Protein), Protein plus added tryptophan at a Protein:Trp molar ratio of 1:0.25 (Protein+Trp~0.25), and Protein plus added tryptophan at a Protein:Trp molar ratio of 1:1 (Protein+Trp~1.0). Table~\ref{tab:qy_tud_both} summarizes TuD quantum yields. For MTs, the microplate absorbance contains a wavelength-dependent scattering contribution that is negligible for TuD but substantial for MT, which complicates estimation of the absorbed fraction at the excitation wavelength (most strongly near 295~nm). We therefore report MT quantum yields as bounds in Table~\ref{tab:qy_mt_bounds_both} using two absorbance treatments: $QY_{\mathrm{raw}}$ (blank-subtracted absorbance; lower-bound yield) and $QY_{\mathrm{bgA}}$ (background-subtracted absorbance after tail fitting; upper-bound yield). We provide the absorbance and integrated fluorescence values used in the yield calculations in Supplementary Tables~S1--S3, and we describe the tail-fitting procedure and fit-quality summary in Supplementary Text (Supplementary Fig.~S1; Supplementary Tables~S6--S7).

\subsubsection*{Polymerization increases yield at 280~nm}
\begin{figure}[t]
	\centering
	\includegraphics[width=0.85\textwidth]{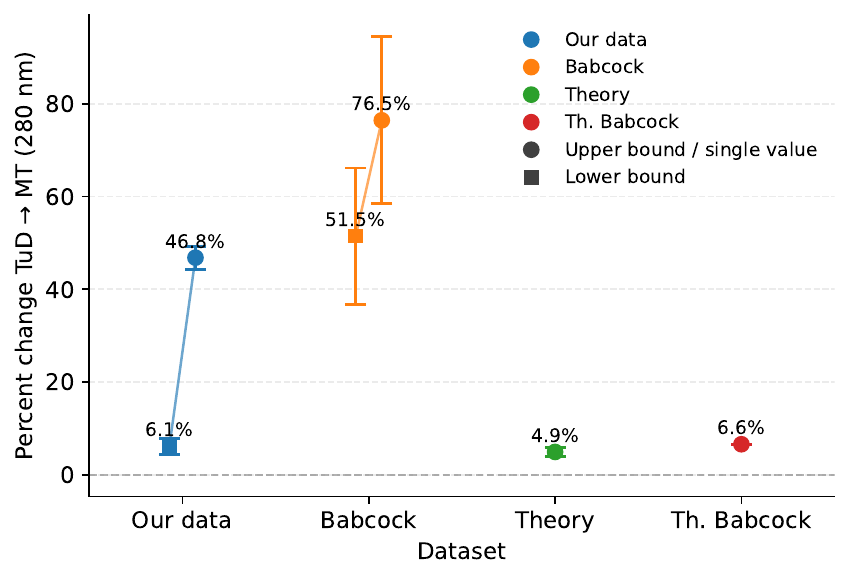}
	\caption{\textbf{Polymerization-associated percent change in quantum yield at 280~nm.}
Percent change in quantum yield upon polymerization (TuD$\rightarrow$MT) at 280~nm.
Experimental values from this work are shown as lower and upper bounds, obtained using $QY_{\mathrm{raw}}$ and $QY_{\mathrm{bgA}}$ for the MT absorbance.
Theory points show predictions from the excitonic model evaluated on molecular-dynamics--derived ensembles.
Literature values from Babcock et al.~\cite{babcock2024ultraviolet} are also shown.
For both our dataset and the Babcock dataset, lower and upper bounds are plotted separately.
Error bars denote propagated experimental uncertainties for the experimental datasets and ensemble variability for theory.
Numerical values are reported in Supplementary Tables~S4--S5.}
	\label{fig:pct_change_tud_mt_280}
\end{figure}

We quantified polymerization-induced changes as the percent change in quantum yield upon polymerization (TuD$\rightarrow$MT) (Fig.~\ref{fig:pct_change_tud_mt_280}). Because MT quantum yields are bounded, the corresponding polymerization percent changes are also bounded. We therefore report the polymerization effect using both MT branches and list the numerical values, with propagated uncertainties, in Supplementary Table~S4. Welch two-sided tests comparing TuD versus MT within each condition are reported in Supplementary Table~S9.

At 280~nm excitation, polymerization produces a clear increase in MT quantum yield relative to TuD, and this conclusion is robust to how MT absorbance is treated. In the no-Trp condition, the polymerization percent change is positive for both bounds, ranging from approximately $+6.1\%$ using $QY_{\mathrm{raw}}$ to approximately $+46.8\%$ using $QY_{\mathrm{bgA}}$ (Fig.~\ref{fig:pct_change_tud_mt_280}; Supplementary Table~S4). Consistent with this, the TuD versus MT difference is statistically significant for both the background-corrected MT estimate ($p=3.018\times10^{-11}$) and the uncorrected estimate ($p=0.003136$) (Supplementary Table~S9). The same branch-resolved framework is applied to the Trp-spiked conditions, with corresponding percent-change bounds reported in Supplementary Table~S4 and statistical tests given in Supplementary Table~S9. Qualitatively, the direction of change, namely enhanced MT emission upon polymerization at 280~nm, is consistent with the trend reported by Babcock et al.~\cite{babcock2024ultraviolet}, although absolute magnitudes are not directly comparable across platforms because scattering systematics, geometry, and analysis choices differ.

\subsubsection*{Polymerization at 295~nm remains bounded}
At 295~nm excitation, the measurement is more selective for tryptophan absorption, but MT absorbance is also more strongly affected by scattering in this geometry. As a result, TuD versus MT comparisons depend sensitively on the MT absorbance branch (Supplementary Table~S9), and the polymerization percent-change bounds span a wide range (Appendix Fig.~\ref{fig:pct_change_tud_mt_295}; Supplementary Table~S5). In the no-Trp condition, the two bounds differ substantially and reverse sign, corresponding to an apparent decrease of about $-58.6\%$ using $QY_{\mathrm{raw}}$ and an apparent increase of about $+12.7\%$ using $QY_{\mathrm{bgA}}$ (Appendix Fig.~\ref{fig:pct_change_tud_mt_295}; Supplementary Table~S5). Because this sign reversal occurs across plausible absorbance treatments, the most reliable conclusion at 295~nm is not a single effect size, but rather that the polymerization comparison remains bounded in this dataset. Tighter control of scattering and inner-filter systematics will be required before assigning a definitive polymerization effect magnitude at tryptophan-selective excitation.

From a theoretical perspective, the expected polymerization-associated enhancement in this parameter regime is modest. In idealized ordered-lattice pictures, collective radiative effects yield only a small enhancement relative to dimers \cite{babcock2024ultraviolet}, and incorporating structural disorder, as in the molecular-dynamics--derived ensembles considered here, typically suppresses the enhancement while preserving the qualitative direction. Consistent with this scale, our excitonic model predicts a polymerization-associated change of order $\sim 5\%$ (mean across $n=20$ realizations; Supplementary Table~S10; baseline: $4.9\%\pm1.0\%$). Given the experimental bounds at 295~nm and the simplified nature of the model, the most meaningful comparison is therefore the trend, namely enhancement versus reduction, rather than the exact magnitude of the percent change.

\subsubsection*{Added tryptophan quenches microtubules}
To quantify tryptophan-dependent changes within MTs, we computed the percent change in MT quantum yield relative to the no-added-Trp MT condition. Here, Trp~0.25 denotes a Protein:Trp molar ratio of 1:0.25, and Trp~1.0 denotes a Protein:Trp molar ratio of 1:1. Because this comparison is performed within MT samples, it is less sensitive to the absolute absorbance baseline than the TuD$\rightarrow$MT polymerization ratio, particularly at 295~nm where scattering dominates the absorbed-fraction estimate. Table~\ref{tab:delta_mt} reports MT percent changes using $QY_{\mathrm{bgA}}$, and Supplementary Table~S8 reports Welch tests for MT condition-to-condition comparisons (for both $QY_{\mathrm{bgA}}$ and $QY_{\mathrm{raw}}$).

Within MTs, added Trp produces a consistent quenching trend at both excitation wavelengths (Table~\ref{tab:delta_mt}). At 280~nm, the MT quantum yield decreases by 7.27\% for Trp~0.25 and by 11.20\% for Trp~1.0. A similar concentration-dependent decrease is observed at 295~nm, with changes of 8.21\% for Trp~0.25 and 16.45\% for Trp~1.0 (Table~\ref{tab:delta_mt}). Branch-resolved Welch tests support these trends: at 280~nm, Protein versus Protein+Trp~0.25 is significant for both absorbance branches, and at 295~nm, Protein versus Protein+Trp~1.0 is significant for both branches, whereas Protein versus Protein+Trp~0.25 is marginal in this dataset (Supplementary Table~S8). For completeness, Supplementary Table~S8 also reports analogous pairwise tests for TuD condition-to-condition comparisons; in contrast to MTs, TuD shows weaker and less systematic Trp-dependent differences across excitation wavelengths in this dataset.

The observed MT quenching is also qualitatively consistent with the excitonic modeling. Using the same perturbations evaluated on molecular-dynamics--derived ensembles, the theory predicts negative MT changes for both Trp conditions (Table~\ref{tab:delta_mt}), supporting agreement in the direction of the effect and motivating future structure-specific and time-resolved measurements to probe the underlying mechanism.

\section*{Discussion}

\subsection*{What the combined evidence supports}

The combined modeling and spectroscopy support three main conclusions that are most naturally interpreted in an engineering
and feasibility framework, namely whether microtubule ultraviolet fluorescence can be predictably tuned by network-scale
effects and by externally applied chemical perturbations under realistic disorder and measurement constraints. First, the
simulations show that structural variability and diagonal energetic disorder substantially suppress the brightest
cooperative radiative channels relative to idealized ordered lattices, placing an upper bound on how large any
superradiance-based enhancement can be in physically realistic assemblies. In this regime, polymerization is predicted
to produce only a modest increase in fluorescence quantum yield relative to isolated dimers, with summary theory values
indicating an enhancement on the order of a few percent in the present parameter range. From a design perspective, this
suggests that large radiative-rate gains are unlikely without additional control of disorder, while modest but systematic
yield shifts remain feasible.

Second, the experiments show that polymerization produces an increase in microtubule quantum yield at 280~nm excitation
that is insensitive to plausible treatments of microtubule absorbance in the microplate geometry. The direction of the
effect is stable under both absorbance branches, and the corresponding TuD versus MT tests remain significant, indicating
that the qualitative polymerization trend at 280~nm is not an artifact of scattering corrections. This observation is
consistent with the theoretical expectation that polymerization can increase quantum yield, while also emphasizing that
the experimentally inferred magnitude depends on how the absorbed fraction is estimated.

Third, 295~nm excitation provides the most direct conceptual connection to a tryptophan-based excitonic model because
excitation is more selective for tryptophan, but it is also the wavelength where microtubule absorbance is most strongly
dominated by scattering in the microplate geometry. For this reason, the TuD to MT polymerization comparison at 295~nm
cannot be assigned a single definitive sign or effect size in the present dataset, and is more appropriately treated as
bounded. Under these conditions, the most reliable experimental constraints at 295~nm come from within-microtubule
comparisons that keep scattering structure and geometry fixed across conditions. These results provide a feasibility
proof-of-principle that a chemical cofactor can shift microtubule quantum yield in a controlled direction. In that
framework, added L-tryptophan produces a reproducible, concentration-dependent reduction in microtubule quantum yield at
both 280 and 295~nm, while the corresponding dimer response is weaker and less systematic. In the engineering context,
this is the clearest evidence in this work that microtubule quantum yield is chemically addressable via an external
cofactor, and that network-level models can anticipate the direction of a tunable response, even when polymerization
effects at tryptophan-selective excitation remain bounded by extinction systematics. Finally, the simulations also
motivate a second realizable perturbation that more directly targets the native network, namely removal of Trp~3 using
tubulin variants that lack the tryptophan~3 site, as occurs in \emph{Trypanosoma brucei}, a parasite associated with
sleeping sickness~\cite{marchese2018uptake}.

\subsection*{Local photophysics versus cooperative emission}

A local, noncooperative picture treats tryptophan residues as largely independent fluorophores connected by incoherent
excitation transfer, often approximated as F\"orster-type hopping. In this limit, energy transfer primarily redistributes
excitation among sites and can introduce additional loss through trapping or quenching, but it does not generically
create new collective radiative channels that increase the ensemble radiative rate or quantum yield. Apparent
polymerization-dependent changes in steady-state yield can therefore arise from conventional sources, including
microenvironment shifts, reabsorption, inner-filter effects, and scattering corrections that couple absorbance estimation
to fluorescence normalization~\cite{Lakowicz2006}. For biotechnology, these local mechanisms are not a drawback per se,
but they must be separated from network-level radiative effects to yield transferable design rules.

Excitonic models admit a different mechanism: polymerization creates a coupled tryptophan network that can support
delocalized eigenstates with enhanced radiative decay relative to independent emitters~\cite{Dicke1954}. In our
framework, structural variability suppresses the brightest collective channels relative to ordered templates, but it does
not preclude systematic differences between microtubules and dimers or between perturbed and unperturbed networks. This
motivates a conservative interpretation in which cooperative radiative contributions, if present, are modest under
realistic disorder and are best evaluated with observables that directly report radiative-rate changes, rather than
inferred solely from steady-state intensities that remain entangled with extinction systematics.

\subsection*{Experimental limitations}

Several experimental approximations limit how precisely absolute quantum yields can be assigned, particularly for
polymerized microtubules at 295~nm excitation. The analysis combines absorbance and fluorescence at the condition-mean
level, which assumes that replicate variability captures plate-to-plate differences and any slow concentration drift.
Small systematic offsets between absorbance and fluorescence measurements could therefore persist without being obvious
from replicate scatter.

In addition, dimers and microtubules were measured in related but not identical chemical environments (BRB80 for dimers;
BRB80 plus taxol for microtubules). Matched blanks mitigate baseline differences and refractive-index effects should be
small in aqueous buffers, but residual mismatch can contribute to uncertainty in absolute scaling. The dominant
limitation for microtubules is scattering in the absorbance readout. In microplate geometry, microtubule extinction
contains a strong wavelength-dependent scattering contribution that becomes most consequential near 295~nm, making
polymerization comparisons at that wavelength intrinsically sensitive to how extinction is decomposed into absorption
versus scattering.

Finally, each well can contain a mixture of polymerized microtubules, unpolymerized tubulin, and aggregates, and the
effective polymer fraction may vary between wells. Added tryptophan introduces an additional ambiguity: free Trp can
remain in solution, occupy binding pockets with unknown occupancy, or associate nonspecifically. Without orthogonal
binding and occupancy assays, tryptophan-dependent changes cannot be uniquely assigned to a specific site or binding mode.

\subsection*{Modeling limitations}

The modeling treats the tryptophan system in the single-excitation manifold using an effective non-Hermitian Hamiltonian,
\(
H_{\mathrm{eff}} = H_{0} + \Delta - \tfrac{i}{2}G,
\)
with radiative couplings computed from dipole--dipole kernels that include near-, intermediate-, and far-field terms. The
model neglects multi-excitation effects, explicit photon modes, and exciton--exciton annihilation. It assigns a common
transition energy and bare radiative rate to all tryptophans, while structural variability enters through
molecular-dynamics conformations and diagonal energetic disorder.

The thermal quantum yield calculation uses Boltzmann-weighted radiative decay rates together with a single phenomenological
nonradiative rate $\Gamma_{\mathrm{nr}}$ that is constant and site independent. This approximation ignores binding-induced
quenching channels and any correlations between local environment, delocalization, and nonradiative decay. Representing
external tryptophan as added or removed dipoles at prescribed locations also omits electronic reorganization and new
nonradiative pathways that can accompany binding, except indirectly through the structural ensembles used.

\subsection*{Next steps}

Future work should prioritize time-resolved measurements that separate radiative and nonradiative rate changes while
improving control of extinction systematics in polymer samples. The steady-state results already indicate that added
L-tryptophan reproducibly suppresses microtubule fluorescence, providing a concrete chemical perturbation that can be
used as a benchmark for engineered tuning. Fluorescence lifetimes and lifetime distributions, anisotropy, temperature
dependence, excitation-power dependence, and polarization-resolved emission would place stronger constraints on whether
cooperative radiative channels contribute beyond conventional photophysical changes. In this context, delayed-luminescence
measurements on tubulin and microtubule constructs provide a complementary time-domain precedent: the reported emission
intensity and decay dynamics depend strongly on assembly state and evolve during polymerization \cite{Zoghi2025BODA}.
Adapting analogous long-time emission observables (alongside conventional lifetimes) within the tryptophan-perturbation
framework could help distinguish genuinely altered radiative channels from predominantly nonradiative quenching.

In parallel, orthogonal assays for cofactor occupancy and binding mode would help translate phenomenological tuning into
site-specific design rules. On the theory side, incorporating heterogeneous nonradiative pathways, vibronic structure,
and contributions from additional aromatic residues will be important for moving from qualitative trend capture toward
quantitative prediction.

\section*{Materials and Methods}
\label{sec:methods}

\subsection*{Experimental design}
This study combines excitonic modeling of tubulin tryptophan networks with steady-state ultraviolet spectroscopy of tubulin dimers (TuD) and taxol-stabilized microtubules (MT). The modeling pipeline constructs microtubule-like assemblies from structural templates and molecular-dynamics (MD) snapshots, assigns tryptophan transition dipoles, and computes collective radiative properties and thermal fluorescence quantum yield using an effective non-Hermitian Hamiltonian. The experimental pipeline measures absorbance and fluorescence in UV-transparent microplates under 280 and 295~nm excitation and converts steady-state signals to absolute quantum yields by comparison to an excitation-matched tryptophan reference. Additional experimental tables, background-fit examples, and complete statistical results are provided in the Supplementary Materials (Fig.~S1; Tables~S1--S10).

\subsection*{Tryptophan network model: Site coordinates and transition dipoles}
The structure of the microtubule is built based on approaches used in earlier studies \cite{celardo2019existence,babcock2024ultraviolet,patwa2024quantum}. Several tubulin dimers, taken from the Protein Data Bank (PDB), are placed next to each other to form a left-handed helix with a diameter of about 22.4~nm. In the starting position, each dimer is arranged so that the $\alpha$- and $\beta$-tubulin chains are aligned along the protofilament direction, which we define as the \( x \)-axis. To model the helical shape, each tubulin dimer goes through a series of steps. First, it is rotated by \(-55.38^\circ\) around its own long axis. Then, it is rotated again by \(11.7^\circ\) around the CD2 atom of the Trp 346 residue in the $\beta$-tubulin (see Fig.~\ref{fig:tub}). After that, the dimer is moved by 11.2~nm in the \( y \)-direction and 0.3~nm in the \( z \)-direction. These steps are repeated to place each new dimer in the helix. For each \(N\)th dimer (with \(N \leq 13\)), we also apply a rotation of \(27.69^\circ\) about the \( x \)-axis and a shift of 0.9~nm along the same axis. This creates a full turn of the microtubule, made up of 13 dimers. To make longer microtubules with several helical turns, we repeat this structure and shift each new turn by 8~nm along the \( x \)-axis. The resulting microtubule has an average radius of about 11.2~nm, measured from the center of the microtubule to the center of mass of the tubulin dimer. Once the microtubule is built, we extract the positions and transition dipole moments of the eight tryptophan (Trp) residues in each tubulin dimer. The position of each Trp is defined as the midpoint between the CD2 and CE2 carbon atoms. The direction of the transition dipole moment is based on the well-known \( ^1L_a \) transition of tryptophan \cite{callis19977}. This dipole is a vector lying in the plane of the indole ring, pointing 46.2$^\circ$ above the line connecting the midpoint of CD2 and CE2 to the CD1 atom, and directed toward the nitrogen atom NE1.

\subsection*{Molecular dynamics simulations and structural ensemble generation}

We performed all-atom molecular dynamics (MD) simulations to quantify structural variability in the tubulin dimer and to generate conformational ensembles for excitonic calculations. We simulated the tubulin dimer in complex with GTP and GDP starting from the Protein Data Bank structure PDB 1TVK. We rebuilt missing residues using SWISS-MODEL \cite{swissmodel} and assigned protonation states at pH~7 with H++ \cite{hpp3}, with histidine tautomer choices refined using PROPKA \cite{propka3}. We carried out the MD simulations with AMBER (AMBER~22, \texttt{pmemd}) \cite{amber2005}. For system setup, the tubulin--nucleotide complex topology was prepared in \texttt{LEaP} (AmberTools) using the ff14SB protein force field \cite{ff14sb} together with standard Amber libraries for the bound nucleotides. The system was solvated and neutralized in \texttt{LEaP} in a rectangular box of TIP3P water with a 25~\AA\ buffer and counterions were added to neutralize the net charge. We applied periodic boundary conditions and constrained bonds involving hydrogen using SHAKE, which enabled a 2~fs time step. After energy minimization and equilibration, we performed a 6.2~ns production run in the NPT ensemble at 310~K using Berendsen temperature and pressure coupling. We saved coordinates every 1{,}000 steps (2~ps), yielding 3{,}100 frames. We aligned trajectory frames and extracted PDB snapshots using \texttt{cpptraj} \cite{cpptraj2013}. We used these 3{,}100 aligned dimer snapshots as structural inputs to assemble microtubule-like lattices, sampling conformations across lattice sites to construct disordered microtubule segments from 1 to 100 spirals (13 dimers per spiral), corresponding to up to 1{,}300 dimers per segment.

To model structural variability under the added-tryptophan condition, we performed an additional MD simulation in which \textit{L}-tryptophan was included explicitly. We prepared this system using the same starting tubulin structure, force field, solvation protocol, and MD settings as in the baseline simulation. After minimization and equilibration, we ran a production trajectory, saved snapshots at the same interval, aligned the frames, and used the resulting ensemble as input to the excitonic calculations for the added-tryptophan scenarios. All simulations were performed on high-performance computing resources provided by the Digital Research Alliance of Canada.

\subsection*{Effective non-Hermitian Hamiltonian and thermal quantum yield}
We modeled collective radiative coupling and decay using an effective non-Hermitian Hamiltonian
\begin{equation}
H_{\mathrm{eff}} = H_{0} + \Delta - \frac{i}{2}G,
\end{equation}
where $H_{0}$ contains site energies, $\Delta$ is the coherent dipole-mediated coupling matrix, and $G$ is the collective radiative decay matrix. We constructed $\Delta$ and $G$ from inter-site separations and dipole orientations using dipole--dipole kernels that include near-, intermediate-, and far-field terms through $\alpha_{nm}=k_{0}r_{nm}$ and the corresponding angular factors, as in prior work~\cite{celardo2019existence,babcock2024ultraviolet,patwa2024quantum}. We used baseline tryptophan parameters (transition wavelength near 280~nm, transition dipole magnitude $\sim 6$~D, and single-site radiative rate $\gamma$) consistent with these references. Explicit expressions for $\Delta_{nm}$ and $G_{nm}$, along with the definitions of $\alpha_{nm}$ and the associated angular factors, are provided in the Supplementary Material.

Diagonalization yields complex eigenvalues $\mathcal{E}_{j}=E_{j}-\tfrac{i}{2}\Gamma_{j}$, where $\Gamma_{j}$ is the radiative decay rate of excitonic state $j$. To connect to steady-state measurements, we computed a Boltzmann-weighted radiative decay rate
\begin{equation}
\langle \Gamma \rangle_{\mathrm{th}} = \frac{1}{Z}\sum_{j}\Gamma_{j}e^{-\beta E_{j}},
\qquad
Z=\sum_{j}e^{-\beta E_{j}},
\end{equation}
and defined the thermal fluorescence quantum yield using a phenomenological nonradiative rate $\Gamma_{\mathrm{nr}}$,
\begin{equation}
\langle \mathrm{QY} \rangle_{\mathrm{th}}=
\frac{\langle \Gamma \rangle_{\mathrm{th}}}{\langle \Gamma \rangle_{\mathrm{th}}+\Gamma_{\mathrm{nr}}}.
\end{equation}
We included static energetic disorder by adding independent site-energy shifts sampled uniformly from $[-W/2,\,W/2]$ (cm$^{-1}$) to $H_{0}$ and averaging results over disorder realizations.

\subsection*{Simulation-guided perturbations}

To generate simulation-guided experimental signatures, we evaluated two classes of perturbations:

(i) \emph{Suppressed tryptophan (proposal):} Trp~3 was removed in silico by deleting the corresponding site from each dimer prior to assembling the MT-like lattice. The resulting ensembles were analyzed using the same Hamiltonian workflow as baseline structures, and the predicted change in $\langle \mathrm{QY}\rangle_{\mathrm{th}}$ was used to motivate a targeted experimental comparison between unmodified tubulin and a Trp~3-suppressed variant (e.g., a Trp~3-lacking tubulin as found in \emph{Trypanosoma brucei}~\cite{marchese2018uptake}).

(ii) \emph{Added tryptophan (simulation and experiment):} Following Yousefzadeh \emph{et al.}~\cite{yousefzadeh2020tryptophan}, an additional tryptophan-like dipole was introduced in simulations near a candidate pocket in $\alpha$-tubulin. We considered (a) unmodified MT-like assemblies, (b) fully modified assemblies (all dimers containing the added tryptophan), and (c) mixed populations in which only a fraction of dimers were modified (e.g., 25\%, 50\%, or 75\%). In the experiments, we probe the same conceptual perturbation by adding free \textit{L}-tryptophan to TuD and MT samples at controlled Trp:protein ratios. While free tryptophan is not restricted to a single binding pocket, it samples a distribution of positions around the protein and therefore serves as an experimental proxy for introducing additional tryptophan-like dipoles into the effective excitonic environment.

\subsection*{Protein samples and ultraviolet spectroscopy}
\subsubsection*{Sample preparation and tryptophan spiking}
We prepared tubulin dimers and taxol-stabilized microtubules (porcine brain) in BRB80 buffer (80~mM PIPES, 2~mM MgCl$_2$, 0.5~mM EGTA, pH~6.9). We prepared free \textit{L}-tryptophan stocks in the same buffer. We performed measurements in 96-well UV-transparent microplates (final volume 110~\textmu L per well) using a Tecan Spark multimode plate reader. 

For TuD, we dissolved 1~mg lyophilized protein in 1.000~mL ice-cold BRB80 to prepare a 1.0~mg~mL$^{-1}$ stock and prepared working batches at 0.300~mg~mL$^{-1}$. We clarified each batch by microcentrifugation (14000~g, 10~min, 4~$^\circ$C) and plated the clear supernatant. For MT, we reconstituted a 0.50~mg vial of taxol-stabilized microtubules to 1.0~mg~mL$^{-1}$ in BRB80 with taxol and diluted to 0.300~mg~mL$^{-1}$ while maintaining taxol. To prepare wells, we added a 10~\textmu L spike (tryptophan or matched buffer) to 100~\textmu L of sample, yielding a final in-well protein concentration of 0.273~mg~mL$^{-1}$.

We selected tryptophan spikes to yield final well concentrations of 0.62 and 2.48~\textmu M, corresponding approximately to Trp:protein ratios of 0.25:1 and 1:1 (Trp per tubulin dimer) at 0.273~mg~mL$^{-1}$ protein.

\subsubsection*{Absorbance and fluorescence acquisition}
We recorded absorbance spectra from 260 to 800~nm and extracted absorbance values at 280 and 295~nm. We recorded fluorescence emission spectra from 310 to 450~nm with excitation at 280 or 295~nm and integrated emission over 320--450~nm (280~nm excitation) or 325--450~nm (295~nm excitation). We used matched buffer wells (including the same tryptophan spike, when applicable) as blanks and processed them identically.

We acquired absorbance and fluorescence in kinetic mode for 1~h at room temperature and analyzed the approximately 30~min time point after plating. We used technical replicates with $n=10$ wells per condition for TuD and $n=5$ wells per condition for MT.

\subsection*{Quantum yield estimation}
\subsubsection*{Absorbance processing and scattering bounds for microtubules}
For TuD, we subtracted the mean matched BRB80 blank (containing the same tryptophan spike) from each absorbance spectrum and read the excitation absorbance from the corrected spectrum.

For MT, blank-subtracted absorbance spectra showed a pronounced wavelength-dependent scattering tail. To bracket the absorbed fraction at the excitation wavelength, we computed two absorbance estimates: (i) the uncorrected blank-subtracted absorbance $A^{\mathrm{raw}}(\lambda_{\mathrm{ex}})$ and (ii) a background-corrected estimate $A^{\mathrm{bg}}(\lambda_{\mathrm{ex}})$ obtained by fitting the long-wavelength region of the blank-subtracted spectrum with a Rayleigh-motivated tail model and subtracting the fitted scattering contribution. A detailed description of the background-fitting procedure is provided in the Supplementary Text. These two branches define lower and upper bounds on MT quantum yield. Representative fits and fit summaries are provided in the Supplementary Materials (Fig.~S1; Tables~S4--S7).

\subsubsection*{Fluorescence processing and absolute quantum yield conversion}
For each well, we integrated the emission spectrum over the specified wavelength window to obtain integrated fluorescence $F$ and subtracted the matched blank integrated over the same window. We obtained absolute quantum yields by comparing each sample to \textit{L}-tryptophan measured with identical optical settings and processed through the same pipeline. We used the reference quantum yield $Q_{r}=0.13$ (Absorbance and fluorescence values are provided in the Supplementary Text). With absorbed fraction $a=1-10^{-A}$, the sample quantum yield is
\begin{equation}
Q_{s}(\lambda_{\mathrm{ex}})=Q_{r}
\frac{F_{s}(\lambda_{\mathrm{ex}})}{F_{r}(\lambda_{\mathrm{ex}})}
\frac{a_{r}(\lambda_{\mathrm{ex}})}{a_{s}(\lambda_{\mathrm{ex}})}
\left(\frac{n_{s}}{n_{r}}\right)^{2}.
\end{equation}
Because all samples and the reference were aqueous, we set $n_{s}=n_{r}$. For MT, we report bounds $(QY_{\mathrm{raw}},QY_{\mathrm{bgA}})$ obtained from $A^{\mathrm{raw}}$ and $A^{\mathrm{bg}}$, respectively, using the same fluorescence $F_{s}$. The absorbance and fluorescence values used in these calculations are tabulated in Supplementary Tables~S1--S3.

\subsection*{Statistical analysis}

We computed replicate means and standard errors across wells within each condition. We propagated quantum-yield uncertainties from absorbance and fluorescence standard errors using first-order error propagation (details in the Supplementary Text).
We assessed pairwise comparisons between conditions using two-sided Welch $t$ tests performed separately for each excitation wavelength and sample type. For MT, we evaluated tests on both absorbance branches. The complete set of tests is reported in the Supplementary Materials (Tables~S8--S9). For theory ensembles, we report summary statistics across $n=20$ realizations and associated tests in Supplementary Table~S10.

Full details of the statistical analysis are provided in the Supplementary Text of the Supplementary Materials.

\section*{Acknowledgments}
We acknowledge the Global Water Futures Observatories Nanobiosensors Laboratory at the University of Waterloo for access to the Tecan Spark multimode microplate reader used in this work.

\paragraph*{Funding:}
This research was undertaken in part thanks to funding to T.J.A.C. from the Canada Research Chairs Program (CRC-2022-00204) and the University of Waterloo. L.G. acknowledges support from the University of Waterloo Provost’s Program for Interdisciplinary Postdoctoral Scholars.

\paragraph*{Author contributions:}
Conceptualization: L.G., T.J.A.C.\\
Methodology: L.G., T.J.A.C.\\
Theoretical modeling and simulations: L.G.\\
Experimental investigation: L.G.\\
Data curation and analysis: L.G.\\
Funding acquisition: T.J.A.C.\\
Writing---original draft: L.G.\\
Writing---review \& editing: L.G., T.J.A.C.

\paragraph*{Competing interests:}
Authors declare that they have no competing interests.

\paragraph*{Data and materials availability:}
All data are available in the main text or the supplementary materials. Raw plate-reader exports and the analysis code used for absorbance processing, tail fitting, quantum-yield calculations, uncertainty propagation, and statistical testing are provided as supplementary files (Data~S1 and Code~S1).

\clearpage


\begin{table} 
	\centering
	\caption{\textbf{Tubulin-dimer quantum yields measured at 280 and 295~nm excitation.}
	Values are means $\pm$ SE at approximately 30~min after plating.}
	\label{tab:qy_tud_both}

	\begin{tabular}{lcccc}
		\\
		\hline
		Sample & QY (280~nm) & SE & QY (295~nm) & SE \\
		\hline
		TuD (no Trp)   & 0.103851 & 0.001482 & 0.138666 & 0.006958 \\
		TuD + Trp 0.25 & 0.104731 & 0.001120 & 0.145850 & 0.005823 \\
		TuD + Trp 1.00 & 0.100577 & 0.001067 & 0.141075 & 0.002577 \\
		\hline
	\end{tabular}
\end{table}

\begin{table} 
	\centering
	\caption{\textbf{Microtubule quantum yields measured at 280 and 295~nm excitation reported as bounds.}
	$QY_{\mathrm{raw}}$ uses blank-subtracted absorbance (lower-bound QY). $QY_{\mathrm{bgA}}$ uses background-corrected
	absorbance after tail fitting (upper-bound QY). Values are means $\pm$ SE.}
	\label{tab:qy_mt_bounds_both}

	\resizebox{\textwidth}{!}{%
	\begin{tabular}{lcccccccc}
		\\
		\hline
		 & \multicolumn{4}{c}{280~nm} & \multicolumn{4}{c}{295~nm} \\
		\hline
		Sample &
		$QY_{\mathrm{raw}}$ & SE & $QY_{\mathrm{bgA}}$ & SE &
		$QY_{\mathrm{raw}}$ & SE & $QY_{\mathrm{bgA}}$ & SE \\
		\hline
		MT (no Trp)   & 0.110183 & 0.000931 & 0.152483 & 0.001388 & 0.057427 & 0.001412 & 0.156245 & 0.003840 \\
		MT + Trp 0.25 & 0.102017 & 0.002337 & 0.141393 & 0.003343 & 0.052757 & 0.001493 & 0.143414 & 0.004245 \\
		MT + Trp 1.0  & 0.097232 & 0.004851 & 0.135407 & 0.006792 & 0.048162 & 0.002902 & 0.130546 & 0.008051 \\
		\hline
	\end{tabular}}
\end{table}

\begin{table} 
	\centering
	\caption{\textbf{Tryptophan-dependent changes in microtubule quantum yield.}
	Changes are reported relative to the no-Trp MT condition using $QY_{\mathrm{bgA}}$.
	Percent changes are defined as $\Delta QY(c)=100\,[Q_{\mathrm{MT,bgA}}^{(c)}/Q_{\mathrm{MT,bgA}}^{(0)}-1]$.
	Theory values correspond to the same perturbations evaluated on molecular-dynamics--derived ensembles.}
	\label{tab:delta_mt}

	\resizebox{\textwidth}{!}{%
	\begin{tabular}{lcccccc}
		\\
		\hline
		Condition &
		$\Delta QY$ exp.\ 280~nm (\%) & SE (\%) &
		$\Delta QY$ exp.\ 295~nm (\%) & SE (\%) &
		$\Delta QY$ theory 280~nm (\%) & SE (\%) \\
		\hline
		Trp 0.25 & -7.27  & 2.35 & -8.21  & 3.53 & -1.6 & 0.09 \\
		Trp 1.0  & -11.20 & 4.53 & -16.45 & 5.55 & -8.9 & 0.07 \\
		\hline
	\end{tabular}}
\end{table}


\clearpage 

%
\bibliography{science_template} 
\bibliographystyle{sciencemag}


\subsection*{Supplementary materials}
Supplementary Text\\
Fig. S1\\
Tables S1 to S10\\
Data S1\\
Code S1


\clearpage




\renewcommand{\thefigure}{S\arabic{figure}}
\renewcommand{\thetable}{S\arabic{table}}
\renewcommand{\theequation}{S\arabic{equation}}
\renewcommand{\thepage}{S\arabic{page}}
\setcounter{figure}{0}
\setcounter{table}{0}
\setcounter{equation}{0}
\setcounter{page}{1} 


\begin{center}
\section*{Supplementary Materials for\\ \scititle}

Lea~Gassab$^{\ast}$,
Travis~J.~A.~Craddock$^{\ast}$\\
\small$^\ast$Corresponding authors. Email: lgassab@uwaterloo.ca (L.G.); travis.craddock@uwaterloo.ca (T.J.A.C.)
\end{center}

\subsubsection*{This PDF file includes:}
Supplementary Text\\
Figure S1\\
Figure S2\\
Tables S1 to S10\\

\subsubsection*{Other Supplementary Materials for this manuscript:}
Data S1\\
Code S1\\

\clearpage


\subsection*{Supplementary Text}

\subsubsection*{Overview of supplementary content}
This Supplementary Materials PDF provides the explicit non-Hermitian Hamiltonian expressions used to compute
superradiant enhancement rates and thermal quantum yields, along with the supporting tables and statistical tests used to
compute experimental quantum yields, the excitation-matched reference measurements, and the absorbance background-fitting
procedure used to bracket microtubule absorbance in the microplate geometry. Figure~S1 shows an example background fit.
Tables~S1 to S3 list the absorbance and integrated fluorescence values used for quantum-yield calculations. Tables~S4 and
S5 provide percent change bounds for polymerization comparisons. Tables~S6 and S7 summarize the background-fitting
parameters and a model sensitivity check. Tables~S8 and S9 report Welch tests for key experimental comparisons.
Table~S10 summarizes the corresponding theory statistics.

\subsubsection*{Calculation of quantum yield and superradiant enhancement rates}
\label{sec:si_qy_superrad}

We compute fluorescence quantum yield (QY) and superradiant enhancement rates using an effective non-Hermitian Hamiltonian
formalism for radiatively coupled dipoles, following Refs.~\cite{celardo2019existence,babcock2024ultraviolet,patwa2024quantum}.
This approach captures cooperative radiative decay in extended networks beyond the strict near-field dipole--dipole limit.

\paragraph*{Effective non-Hermitian Hamiltonian.}
The single-excitation manifold is described by
\begin{equation}
H_{\mathrm{eff}} = H_{0} + \Delta - \frac{i}{2}G,
\end{equation}
where \(H_{0}\) contains site energies, \(\Delta\) is the coherent radiative coupling matrix, and \(G\) is the collective
radiative decay matrix. For a network of \(N\) chromophores,
\begin{equation}
H_{0} = \sum_{n=1}^{N} \hbar\omega_{n}\,|n\rangle\langle n|.
\end{equation}
In the baseline (no diagonal disorder) case, \(\omega_n=\omega_0\) for all sites. For tryptophan, we use a representative
transition wavelength \(\lambda_{0}\approx 280~\mathrm{nm}\) (equivalently \(\tilde{\nu}_{0}\approx 3.57\times 10^{4}~\mathrm{cm^{-1}}\)),
a transition dipole magnitude \(\mu \approx 6~\mathrm{D}\), and a single-site radiative rate \(\gamma\) consistent with
Refs.~\cite{babcock2024ultraviolet,patwa2024quantum}.

For \(n\neq m\), define \(r_{nm}=|\mathbf{r}_n-\mathbf{r}_m|\), \(\hat{\mathbf{r}}_{nm}=(\mathbf{r}_n-\mathbf{r}_m)/r_{nm}\),
\(\hat{\boldsymbol{\mu}}_n\) the unit transition dipole at site \(n\), and \(\alpha_{nm}=k_{0}r_{nm}\) with \(k_{0}=2\pi/\lambda_{0}\).
The off-diagonal elements of \(\Delta\) and \(G\) are \cite{celardo2019existence,babcock2024ultraviolet,patwa2024quantum}
\begin{equation}
\Delta = \sum_{n\neq m}\Delta_{nm}\,|n\rangle\langle m|,
\end{equation}
\begin{equation}
\begin{aligned}
\Delta_{nm} &= \frac{3\gamma}{4}\Biggl[
\left(-\frac{\cos\alpha_{nm}}{\alpha_{nm}}+\frac{\sin\alpha_{nm}}{\alpha_{nm}^{2}}+\frac{\cos\alpha_{nm}}{\alpha_{nm}^{3}}\right)
(\hat{\boldsymbol{\mu}}_{n}\!\cdot\!\hat{\boldsymbol{\mu}}_{m}) \\
&\qquad\qquad -
\left(-\frac{\cos\alpha_{nm}}{\alpha_{nm}}+\frac{3\sin\alpha_{nm}}{\alpha_{nm}^{2}}+\frac{3\cos\alpha_{nm}}{\alpha_{nm}^{3}}\right)
(\hat{\boldsymbol{\mu}}_{n}\!\cdot\!\hat{\mathbf{r}}_{nm})(\hat{\boldsymbol{\mu}}_{m}\!\cdot\!\hat{\mathbf{r}}_{nm})
\Biggr],
\end{aligned}
\end{equation}
and
\begin{equation}
G = \sum_{n=1}^{N}\gamma\,|n\rangle\langle n| + \sum_{n\neq m}G_{nm}\,|n\rangle\langle m|,
\end{equation}
\begin{equation}
\begin{aligned}
G_{nm} &= \frac{3\gamma}{2}\Biggl[
\left(\frac{\sin\alpha_{nm}}{\alpha_{nm}}+\frac{\cos\alpha_{nm}}{\alpha_{nm}^{2}}-\frac{\sin\alpha_{nm}}{\alpha_{nm}^{3}}\right)
(\hat{\boldsymbol{\mu}}_{n}\!\cdot\!\hat{\boldsymbol{\mu}}_{m}) \\
&\qquad\qquad -
\left(\frac{\sin\alpha_{nm}}{\alpha_{nm}}+\frac{3\cos\alpha_{nm}}{\alpha_{nm}^{2}}-\frac{3\sin\alpha_{nm}}{\alpha_{nm}^{3}}\right)
(\hat{\boldsymbol{\mu}}_{n}\!\cdot\!\hat{\mathbf{r}}_{nm})(\hat{\boldsymbol{\mu}}_{m}\!\cdot\!\hat{\mathbf{r}}_{nm})
\Biggr].
\end{aligned}
\end{equation}
Diagonalization yields complex eigenvalues
\begin{equation}
\mathcal{E}_{j}=E_{j}-\frac{i}{2}\Gamma_{j},
\end{equation}
where \(E_j\) is the excitonic energy and \(\Gamma_j\) is the radiative decay rate of eigenstate \(j\).

\paragraph*{Superradiant enhancement.}
We quantify superradiance using the enhancement factor
\begin{equation}
\eta_{\mathrm{SR}}=\frac{\Gamma_{\max}}{\gamma},
\qquad
\Gamma_{\max}=\max_{j}\Gamma_{j},
\end{equation}
so that \(\eta_{\mathrm{SR}}>1\) indicates a superradiant state relative to an isolated chromophore with rate \(\gamma\).

\paragraph*{Thermal fluorescence quantum yield.}
To connect the radiative spectrum of \(H_{\mathrm{eff}}\) to steady-state measurements, we compute a Boltzmann-weighted
radiative rate
\begin{equation}
\langle \Gamma \rangle_{\mathrm{th}} = \frac{1}{Z}\sum_{j}\Gamma_{j}e^{-\beta E_{j}},
\qquad
Z=\sum_{j}e^{-\beta E_{j}},
\end{equation}
with \(\beta=(k_{B}T)^{-1}\). The thermal fluorescence quantum yield is then defined using a phenomenological,
state-independent nonradiative rate \(\Gamma_{\mathrm{nr}}\),
\begin{equation}
\langle \mathrm{QY} \rangle_{\mathrm{th}}=
\frac{\langle \Gamma \rangle_{\mathrm{th}}}{\langle \Gamma \rangle_{\mathrm{th}}+\Gamma_{\mathrm{nr}}}.
\end{equation}
Unless otherwise stated, all reported quantum yields refer to this thermal quantum yield.

\subsubsection*{Absorbance and fluorescence values used in quantum-yield calculations}
Tables~S1 to S3 report the mean absorbance at the excitation wavelength and the integrated fluorescence used in the
quantum-yield calculations. For tubulin dimers, absorbance values are obtained by buffer blank subtraction only. For
microtubules, the blank-subtracted absorbance contains a strong scattering contribution in the ultraviolet and visible.
Therefore, for microtubules we report both the uncorrected blank-subtracted absorbance and a background-corrected
estimate obtained by fitting the long-wavelength tail to a Rayleigh-motivated baseline and subtracting only the
wavelength-dependent term. Figure~S1 provides an example of this correction workflow. Fit parameter summaries are given
in Table~S6.

\subsubsection*{Excitation-matched tryptophan reference measurements}
The tryptophan reference was measured separately at each excitation wavelength using the same instrument settings and
the same integration window as for protein samples. The resulting reference values used in the quantum-yield conversion
are provided below and are used throughout the main paper and this Supplementary Materials PDF.
\[
A_{280}^{(\mathrm{Trp})}=0.31554 \pm 0.000968814,\qquad
F_{280}^{(\mathrm{Trp})}=(1.512729\times 10^{6}) \pm (5.925774832\times 10^{3})
\]
\[
A_{295}^{(\mathrm{Trp})}=0.07552 \pm 0.000294958,\qquad
F_{295}^{(\mathrm{Trp})}=(9.596739\times 10^{5}) \pm (8.616350537\times 10^{3})
\]
Under the present aqueous conditions we approximate the refractive-index correction factor as unity.

\subsubsection*{Polymerization percent change bounds}
To quantify polymerization-induced changes while retaining the microtubule scattering uncertainty, we report percent
change using both microtubule quantum-yield bounds. Table~S4 reports bounds at 280~nm excitation and Table~S5 reports
bounds at 295~nm excitation.
\[
\%\Delta_{\mathrm{raw}} = 100\left(\frac{QY_{\mathrm{MT,raw}}}{QY_{\mathrm{TuD}}}-1\right),\qquad
\%\Delta_{\mathrm{bgA}} = 100\left(\frac{QY_{\mathrm{MT,bgA}}}{QY_{\mathrm{TuD}}}-1\right)
\]

\subsubsection*{Rayleigh-tail background fitting for microtubule absorbance}
In UV--Vis microplate measurements, the reported absorbance is an apparent extinction that can include both true
molecular absorption and elastic scattering losses that remove light from the detection path. For filamentous assemblies
such as taxol-stabilized microtubules, scattering can produce a smooth long-wavelength tail that extends beyond the
intrinsic aromatic absorption band, consistent with turbidity-style extinction readouts used in microtubule preparations
\cite{gaskin1974turbidimetric}. Because the quantum-yield conversion depends on the absorbed fraction
\(a = 1 - 10^{-A}\), uncertainty in the excitation-wavelength absorbance introduces systematic uncertainty in the
absolute microtubule quantum yield.

We compute a blank-subtracted spectrum for each microtubule well
\[
A_{\mathrm{bs}}(\lambda) \;=\; A_{\mathrm{well}}(\lambda) \;-\; \langle A_{\mathrm{blank}}(\lambda)\rangle,
\]
where the blank is the matched BRB80+taxol group mean, including the same tryptophan spike when present. We then fit the
long-wavelength region from 307 to 800~nm of \(A_{\mathrm{bs}}(\lambda)\) to a Rayleigh-motivated power law plus a constant
offset,
\begin{equation}
A_{\mathrm{fit}}(\lambda)= C + A_{0}\left(\frac{\lambda_{0}}{\lambda}\right)^{4},
\label{eq:rayleigh_fit_sup}
\end{equation}
with \(\lambda_{0}=400\)~nm. The \(\lambda^{-4}\) dependence is motivated by the Rayleigh limit, for which scattering
cross section and extinction scale approximately as \(\lambda^{-4}\) \cite{bohren1983absorption,vandehulst1957light}.
The offset term \(C\) captures any residual wavelength-independent baseline remaining after blank subtraction. In the
correction step, we do not remove \(C\). Instead, we remove only the wavelength-dependent term to avoid over-correcting
the near-UV region.

The background-corrected spectrum is obtained by
\[
A_{\mathrm{corr}}(\lambda)= A_{\mathrm{bs}}(\lambda) - A_{0}\left(\frac{\lambda_{0}}{\lambda}\right)^{4},
\]
and the excitation-wavelength corrected absorbance used for quantum-yield estimation is
\(A^{\mathrm{bg}}(\lambda_{\mathrm{ex}})=A_{\mathrm{corr}}(\lambda_{\mathrm{ex}})\).
Figure~S1 shows an example fit and correction. Table~S6 summarizes fit parameters across conditions.

As a sensitivity check, we evaluated alternative tail models on the same fit window: a free-exponent power law
\(C + A_{0}(\lambda_{0}/\lambda)^{\alpha}\) and a two-term mixture
\(C + A_{4}(\lambda_{0}/\lambda)^{4} + A_{2}(\lambda_{0}/\lambda)^{2}\), as commonly discussed in the context of
\AA ngstr\"om-type spectral exponents \cite{schuster2006angstrom}. In this dataset, these alternatives change the
background-corrected absorbance at the excitation wavelengths by only a few percent and therefore shift the MT upper-bound
quantum yield by only a few percent when fluorescence is held fixed (Table~S7).

\subsubsection*{Uncertainty estimation and error propagation}
Absorbance and fluorescence were treated consistently at the replicate-well level. For each replicate well, one value
was extracted: integrated fluorescence emission \(F\) or absorbance at the excitation wavelength \(A\). For each group
(sample or matched blank; same buffer and Trp spike), we computed the replicate mean \(\bar X\), the replicate standard
deviation \(s_X\), and the standard error \(\mathrm{SE}(\bar X)=s_X/\sqrt{n}\). Blank subtraction was applied to group
means,
\[
\Delta\bar X=\bar X_{\mathrm{sample}}-\bar X_{\mathrm{blank}},
\]
and the uncertainty on \(\Delta\bar X\) is the propagated standard error of this difference,
\[
\mathrm{SE}(\Delta\bar X)=
\sqrt{\mathrm{SE}(\bar X_{\mathrm{sample}})^2+\mathrm{SE}(\bar X_{\mathrm{blank}})^2}
=
\sqrt{\frac{s_{X,\mathrm{sample}}^{2}}{n_{\mathrm{sample}}}+
      \frac{s_{X,\mathrm{blank}}^{2}}{n_{\mathrm{blank}}}}.
\]

Quantum yields were computed from
\[
Q_{\mathrm{s}} = Q_{\mathrm{r}}
\frac{F_{\mathrm{s}}}{F_{\mathrm{r}}}
\frac{a_{\mathrm{r}}}{a_{\mathrm{s}}}
\left(\frac{n_{\mathrm{s}}}{n_{\mathrm{r}}}\right)^{2},
\qquad
a = 1-10^{-A}.
\]
The absorbance standard error was mapped to absorbed-fraction standard error by first-order propagation,
\[
\mathrm{SE}(a)=\left|\frac{da}{dA}\right|\mathrm{SE}(A)=\ln(10)\,10^{-A}\,\mathrm{SE}(A).
\]
Treating \(Q_{\mathrm{r}}\) and \((n_{\mathrm{s}}/n_{\mathrm{r}})^2\) as constants, the relative standard error on the
sample quantum yield was obtained by quadrature of independent relative standard errors,
\[
\left(\frac{\mathrm{SE}(Q_{\mathrm{s}})}{Q_{\mathrm{s}}}\right)^{2}
=
\left(\frac{\mathrm{SE}(F_{\mathrm{s}})}{F_{\mathrm{s}}}\right)^{2}
+
\left(\frac{\mathrm{SE}(F_{\mathrm{r}})}{F_{\mathrm{r}}}\right)^{2}
+
\left(\frac{\mathrm{SE}(a_{\mathrm{s}})}{a_{\mathrm{s}}}\right)^{2}
+
\left(\frac{\mathrm{SE}(a_{\mathrm{r}})}{a_{\mathrm{r}}}\right)^{2}.
\]
Throughout, values are reported as mean \(\pm\) SE, where SE denotes the standard error of the mean across replicate wells.

\subsubsection*{Statistical tests for experimental condition comparisons}
Pairwise differences between tryptophan conditions were assessed using two sided Welch $t$ tests, performed separately for
each excitation wavelength (280 and 295~nm) and sample type (tubulin dimers, TuD, or microtubules, MT). All statistical
calculations were carried out in Python using SciPy. When replicate well level values were available, we used the
two sample unequal variance $t$ test implemented in SciPy (ttest\_ind with equal\_var=False). When only condition level
summaries were used, namely the mean quantum yield $\bar{Q}_i$, its standard error $\mathrm{SE}_i$, and the replicate
count $n_i$, the same Welch test can be written directly in terms of standard errors as
\[
t \;=\; \frac{\bar{Q}_1-\bar{Q}_2}{\sqrt{\mathrm{SE}_1^{2}+\mathrm{SE}_2^{2}}},
\]
with effective degrees of freedom given by the Welch Satterthwaite approximation
\[
\nu \;=\;
\frac{\left(\mathrm{SE}_1^{2}+\mathrm{SE}_2^{2}\right)^{2}}
{\frac{\mathrm{SE}_1^{4}}{n_1-1}+\frac{\mathrm{SE}_2^{4}}{n_2-1}}.
\]
The two sided $p$ value was obtained from the $t$ distribution with $\nu$ degrees of freedom. For the
experimental plates, replicate counts are TuD: $n=10$ and MT: $n=5$. Results are labeled Significant for $p<0.05$,
Marginal for $0.05 \le p < 0.06$, and Not significant otherwise.

Results are reported in Table~S8 for within-sample comparisons and in
Table~S9 for tubulin dimer versus microtubule comparisons within each matched condition.

\subsubsection*{Theory summary statistics}
Table~S10 summarizes the theoretical tubulin and microtubule quantum yields and polymerization-induced percent change
under each condition, together with the corresponding Welch tests used in the main text.

\clearpage


\begin{figure} 
	\centering
	\includegraphics[width=0.85\textwidth]{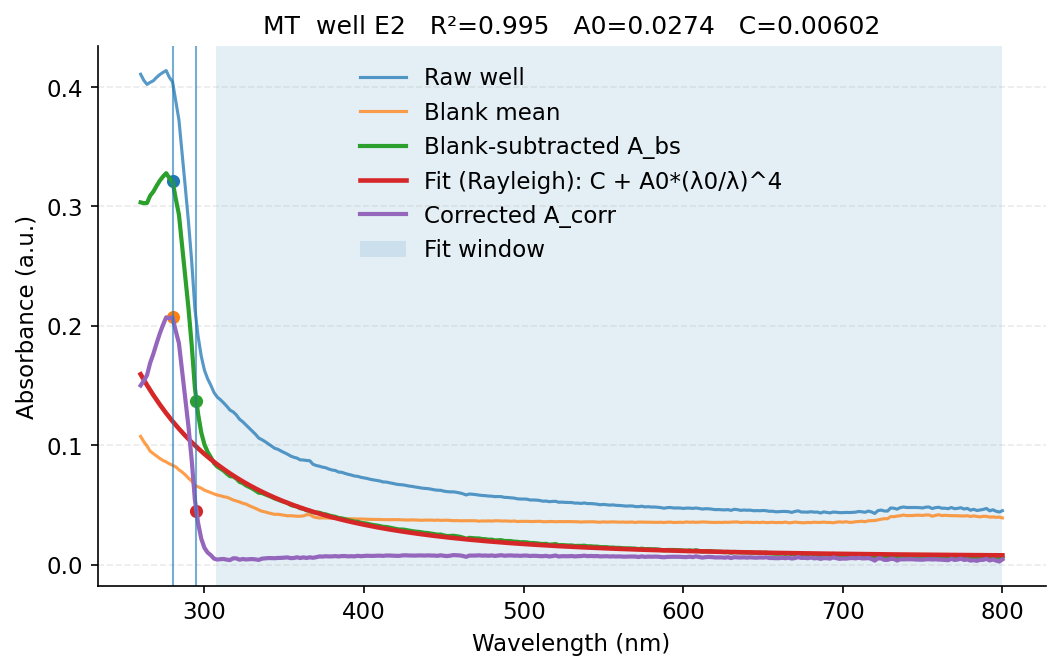}
	\caption{\textbf{Background fitting and scattering correction for microtubule absorbance.}
	Example of microtubule absorbance background fitting for a single well. The Rayleigh-motivated model was fit to the
	blank-subtracted spectrum over 307--800~nm. The fitted long-wavelength tail and the corrected spectrum are overlaid
	for visual quality control of the fit and the resulting correction.}
	\label{fig:sup_bgfit}
\end{figure}

\clearpage

\begin{figure}[t]
	\centering
	\includegraphics[width=0.85\textwidth]{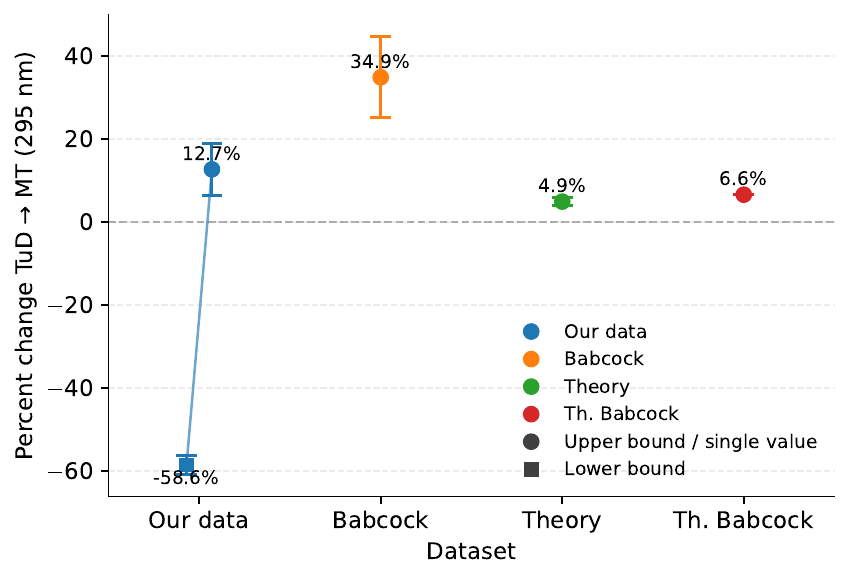}
	\caption{\textbf{Polymerization-associated percent change in quantum yield at 295~nm.}
Percent change in quantum yield upon polymerization (TuD$\rightarrow$MT) at 295~nm.
Experimental values from this work are shown as lower and upper bounds, obtained using $QY_{\mathrm{raw}}$ and $QY_{\mathrm{bgA}}$ for the MT absorbance.
Theory points show predictions from the excitonic model evaluated on molecular-dynamics--derived ensembles.
Literature values labeled Babcock are reproduced from~\cite{babcock2024ultraviolet}.
At 295~nm, the Babcock dataset is represented by a single reported value, whereas our dataset retains distinct lower and upper bounds.
Error bars denote propagated experimental uncertainties for the experimental datasets and ensemble variability for theory.
Numerical values are reported in Supplementary Tables~S4--S5.}
	\label{fig:pct_change_tud_mt_295}
\end{figure}
\clearpage


\begin{table} 
	\centering
	\caption{\textbf{Tubulin dimer absorbance and fluorescence at 280 and 295~nm excitation.}
	Values correspond to approximately 30~min after plating. $A_{\lambda}$ is the blank-subtracted absorbance at the
	excitation wavelength. $F_{\lambda}$ is the blank-subtracted integrated emission (instrument units) over the
	analysis windows used for the quantum-yield calculations.}
	\label{tab:sup_tud_AF}

	\resizebox{\textwidth}{!}{%
	\begin{tabular}{lcccccccc}
		\\
		\hline
		Sample &
		$A_{280}$ & SE($A_{280}$) & $F_{280}$ & SE($F_{280}$) &
		$A_{295}$ & SE($A_{295}$) & $F_{295}$ & SE($F_{295}$)\\
		& & & (a.u.) & (a.u.) & & & (a.u.) & (a.u.)\\
		\hline
		TuD (no Trp)   & 0.28773 & 0.003069 & 1\,133\,621.9 & 12\,792.460173 & 0.05354 & 0.002523 & 743\,849.8 & 16\,032.809098 \\
		TuD + Trp 0.25 & 0.28506 & 0.002477 & 1\,135\,721.6 &  8\,568.558156 & 0.04974 & 0.001650 & 729\,980.2 & 16\,653.691014 \\
		TuD + Trp 1.0  & 0.28024 & 0.001627 & 1\,077\,558.9 &  9\,380.730048 & 0.04758 & 0.000598 & 677\,070.8 &  6\,722.832231 \\
		\hline
	\end{tabular}}
\end{table}

\clearpage

\begin{table} 
	\centering
	\caption{\textbf{Microtubule absorbance and fluorescence at 280~nm excitation.}
	$A_{280}^{\mathrm{raw}}$ is the blank-subtracted absorbance without scattering correction.
	$A_{280}^{\mathrm{bg}}$ is the background-corrected estimate after Rayleigh tail fitting.
	$F_{280}$ is the blank-subtracted integrated emission (instrument units) over the 280~nm analysis window.}
	\label{tab:sup_mt_280_AF}

	\begin{tabular}{lcccccc}
		\\
		\hline
		Sample &
		$A_{280}^{\mathrm{raw}}$ & SE &
		$A_{280}^{\mathrm{bg}}$ & SE &
		$F_{280}$ & SE($F_{280}$)\\
		& & & & & (a.u.) & (a.u.)\\
		\hline
		MT (no Trp)    & 0.31086 & 0.002785 & 0.200234 & 0.001772 & 1\,269\,124.6 &  4\,736.115577 \\
		MT + Trp 0.25  & 0.32666 & 0.005942 & 0.208611 & 0.003270 & 1\,215\,204.6 & 22\,463.544028 \\
		MT + Trp 1.0   & 0.32900 & 0.007303 & 0.208611 & 0.004092 & 1\,163\,758.4 & 55\,116.130514 \\
		\hline
	\end{tabular}
\end{table}

\clearpage

\begin{table} 
	\centering
	\caption{\textbf{Microtubule absorbance and fluorescence at 295~nm excitation.}
	$A_{295}^{\mathrm{raw}}$ is the blank-subtracted absorbance without scattering correction.
	$A_{295}^{\mathrm{bg}}$ is the background-corrected estimate after Rayleigh tail fitting.
	$F_{295}$ is the blank-subtracted integrated emission (instrument units) over the 295~nm analysis window.}
	\label{tab:sup_mt_295_AF}

	\begin{tabular}{lcccccc}
		\\
		\hline
		Sample &
		$A_{295}^{\mathrm{raw}}$ & SE &
		$A_{295}^{\mathrm{bg}}$ & SE &
		$F_{295}$ & SE($F_{295}$)\\
		& & & & & (a.u.) & (a.u.)\\
		\hline
		MT (no Trp)    & 0.13455 & 0.001009 & 0.044755 & 0.000246 & 707\,610.6 & 15\,328.818620 \\
		MT + Trp 0.25  & 0.14318 & 0.003152 & 0.047360 & 0.000981 & 685\,280.8 & 12\,978.410501 \\
		MT + Trp 1.0   & 0.14608 & 0.003884 & 0.048361 & 0.001301 & 636\,260.2 & 35\,047.877509 \\
		\hline
	\end{tabular}
\end{table}

\clearpage

\begin{table} 
	\centering
	\caption{\textbf{Polymerization-induced percent change bounds at 280~nm excitation (tubulin dimer to microtubule).}
	$QY_{\mathrm{MT,raw}}$ and $QY_{\mathrm{MT,bgA}}$ are microtubule quantum-yield bounds obtained from raw and
	background-corrected absorbance branches, respectively. Percent change is reported as
	$100(QY_{\mathrm{MT}}/QY_{\mathrm{TuD}}-1)$.}
	\label{tab:sup_pct_280}

	\resizebox{\textwidth}{!}{%
	\begin{tabular}{lcccccccc}
		\\
		\hline
		Sample (280) &
		QY TuD & SE &
		$QY_{\mathrm{MT,raw}}$ & SE &
		$QY_{\mathrm{MT,bgA}}$ & SE &
		$\%\Delta_{\mathrm{raw}}$ / $\%\Delta_{\mathrm{bgA}}$\\
		\hline
		Protein (no Trp)            & 0.103851 & 0.001482 & 0.110183 & 0.000931 & 0.152483 & 0.001388 & +6.10\% / +46.83\% \\
		Protein + Trp 0.25 (0.25:1) & 0.104731 & 0.001120 & 0.102017 & 0.002337 & 0.141393 & 0.003343 & -2.59\% / +35.01\% \\
		Protein + Trp 1.0 (1:1)     & 0.100577 & 0.001067 & 0.097232 & 0.004851 & 0.135407 & 0.006792 & -3.33\% / +34.63\% \\
		\hline
	\end{tabular}}
\end{table}

\clearpage

\begin{table} 
	\centering
	\caption{\textbf{Polymerization-induced percent change bounds at 295~nm excitation (tubulin dimer to microtubule).}
	$QY_{\mathrm{MT,raw}}$ and $QY_{\mathrm{MT,bgA}}$ are microtubule quantum-yield bounds obtained from raw and
	background-corrected absorbance branches, respectively. Percent change is reported as
	$100(QY_{\mathrm{MT}}/QY_{\mathrm{TuD}}-1)$.}
	\label{tab:sup_pct_295}

	\resizebox{\textwidth}{!}{%
	\begin{tabular}{lcccccccc}
		\\
		\hline
		Sample (295) &
		QY TuD & SE &
		$QY_{\mathrm{MT,raw}}$ & SE &
		$QY_{\mathrm{MT,bgA}}$ & SE &
		$\%\Delta_{\mathrm{raw}}$ / $\%\Delta_{\mathrm{bgA}}$\\
		\hline
		Protein (no Trp)            & 0.138666 & 0.006958 & 0.057427 & 0.001412 & 0.156245 & 0.003840 & -58.59\% / +12.68\% \\
		Protein + Trp 0.25 (0.25:1) & 0.145850 & 0.005823 & 0.052757 & 0.001493 & 0.143414 & 0.004245 & -63.83\% / -1.67\% \\
		Protein + Trp 1.0 (1:1)     & 0.141075 & 0.002577 & 0.048162 & 0.002902 & 0.130546 & 0.008051 & -65.86\% / -7.46\% \\
		\hline
	\end{tabular}}
\end{table}

\clearpage

\begin{table} 
	\centering
	\caption{\textbf{Summary of Rayleigh-tail fit parameters for microtubule blank-subtracted absorbance.}
	Fit window: 307--800~nm with $\lambda_{0}=400$~nm in Eq.~\ref{eq:rayleigh_fit_sup}. Values are mean $\pm$ SE across
	wells within each condition (microtubules, $n=5$).}
	\label{tab:sup_fit_params}

	\resizebox{\textwidth}{!}{%
	\begin{tabular}{lcccccccccc}
		\\
		\hline
		Group & $n$ & $C_{\mathrm{mean}}$ & $C_{\mathrm{SE}}$ & $A0_{\mathrm{mean}}$ & $A0_{\mathrm{SE}}$ &
		$R2_{\mathrm{mean}}$ & $R2_{\mathrm{SE}}$ & $\mathrm{RMSE}_{\mathrm{mean}}$ & $\mathrm{RMSE}_{\mathrm{SE}}$\\
		\hline
		MT            & 5 & 0.006123 & 0.000214 & 0.026561 & 0.000286 & 0.994632 & 0.000159 & 0.001262 & 0.000022 \\
		MT + Trp 0.25 & 5 & 0.005964 & 0.000242 & 0.028344 & 0.000678 & 0.994859 & 0.000214 & 0.001316 & 0.000029 \\
		MT + Trp 1.00 & 5 & 0.006722 & 0.000201 & 0.028905 & 0.000780 & 0.994812 & 0.000212 & 0.001351 & 0.000049 \\
		\hline
	\end{tabular}}
\end{table}

\clearpage

\begin{table} 
	\centering
	\caption{\textbf{Sensitivity of background-corrected absorbance to tail-model choice.}
	Fit window: 307--800~nm with $\lambda_{0}=400$~nm and $n=5$ wells per condition. Alternative models were (i) a
	free-exponent power law \(C+A_{0}(\lambda_{0}/\lambda)^{\alpha}\) and (ii) a two-term mixture
	\(C+A_{4}(\lambda_{0}/\lambda)^{4}+A_{2}(\lambda_{0}/\lambda)^{2}\) \cite{schuster2006angstrom}. Percent changes are
	reported relative to the fixed-exponent Rayleigh model.}
	\label{tab:sup_model_sens}

	\small
	\setlength{\tabcolsep}{5pt}
	\renewcommand{\arraystretch}{1.15}

	\begin{tabular}{lcc|cccccc}
		\hline
		\multicolumn{9}{c}{280~nm} \\
		\hline
		Condition & $n$ & $\alpha$ free &
		$A^{\mathrm{bg}}_{\mathrm{Ray}}$ &
		$A^{\mathrm{bg}}_{\mathrm{free}}$ &
		$A^{\mathrm{bg}}_{\mathrm{2term}}$ &
		$\Delta A$ free (\%) &
		$\Delta QY$ free (\%) &
		$\Delta QY$ 2term (\%) \\
		\hline
		MT            & 5 & 3.37 & 0.200234 & 0.208312 & 0.204641 & +4.03 & -3.05 & -1.69 \\
		MT + Trp 0.25 & 5 & 3.37 & 0.208611 & 0.217235 & 0.213230 & +4.13 & -3.09 & -1.69 \\
		MT + Trp 1.00 & 5 & 3.36 & 0.208611 & 0.217537 & 0.213397 & +4.28 & -3.19 & -1.75 \\
		\hline
		\multicolumn{9}{c}{295~nm} \\
		\hline
		Condition & $n$ & $\alpha$ free &
		$A^{\mathrm{bg}}_{\mathrm{Ray}}$ &
		$A^{\mathrm{bg}}_{\mathrm{free}}$ &
		$A^{\mathrm{bg}}_{\mathrm{2term}}$ &
		$\Delta A$ free (\%) &
		$\Delta QY$ free (\%) &
		$\Delta QY$ 2term (\%) \\
		\hline
		MT            & 5 & 3.37 & 0.044755 & 0.048533 & 0.045741 & +8.44 & -7.39 & -2.05 \\
		MT + Trp 0.25 & 5 & 3.37 & 0.047360 & 0.051397 & 0.048393 & +8.52 & -7.43 & -2.02 \\
		MT + Trp 1.00 & 5 & 3.36 & 0.048361 & 0.052533 & 0.049432 & +8.63 & -7.51 & -2.05 \\
		\hline
	\end{tabular}
\end{table}

\clearpage

\begin{table} 
	\centering
	\caption{\textbf{Welch tests for quantum-yield differences between tryptophan conditions.}
	Two-sided Welch tests were performed separately for each excitation wavelength and sample type. For microtubules,
	tests are reported for both absorbance branches (raw and bgA).}
	\label{tab:sup_welch_conditions}

	\resizebox{\textwidth}{!}{%
	\begin{tabular}{llll}
		\\
		\hline
		Sample type and excitation & Comparison & $p$ value & Interpretation \\
		\hline
		TuD 280 & Protein vs Protein + Trp 0.25 & 0.641810 & Not significant \\
		TuD 280 & Protein vs Protein + Trp 1.0  & 0.091512 & Not significant \\
		TuD 280 & Protein + Trp 0.25 vs Protein + Trp 1.0 & 0.015135 & Significant \\
		TuD 295 & Protein vs Protein + Trp 0.25 & 0.439109 & Not significant \\
		TuD 295 & Protein vs Protein + Trp 1.0  & 0.751300 & Not significant \\
		TuD 295 & Protein + Trp 0.25 vs Protein + Trp 1.0 & 0.467327 & Not significant \\
		\hline
		MT 280 (bgA) & Protein vs Protein + Trp 0.25 & 0.0256925 & Significant \\
		MT 280 (bgA) & Protein vs Protein + Trp 1.0  & 0.0646266 & Not significant \\
		MT 280 (bgA) & Protein + Trp 0.25 vs Protein + Trp 1.0 & 0.460040 & Not significant \\
		MT 295 (bgA) & Protein vs Protein + Trp 0.25 & 0.0556137 & Marginal \\
		MT 295 (bgA) & Protein vs Protein + Trp 1.0  & 0.0295175 & Significant \\
		MT 295 (bgA) & Protein + Trp 0.25 vs Protein + Trp 1.0 & 0.206636 & Not significant \\
		\hline
		MT 280 (raw) & Protein vs Protein + Trp 0.25 & 0.0213134 & Significant \\
		MT 280 (raw) & Protein vs Protein + Trp 1.0  & 0.0546273 & Marginal \\
		MT 280 (raw) & Protein + Trp 0.25 vs Protein + Trp 1.0 & 0.409754 & Not significant \\
		MT 295 (raw) & Protein vs Protein + Trp 0.25 & 0.0527778 & Marginal \\
		MT 295 (raw) & Protein vs Protein + Trp 1.0  & 0.0295376 & Significant \\
		MT 295 (raw) & Protein + Trp 0.25 vs Protein + Trp 1.0 & 0.208948 & Not significant \\
		\hline
	\end{tabular}}
\end{table}

\clearpage

\begin{table} 
	\centering
	\caption{\textbf{Welch tests comparing tubulin dimers versus microtubules within each condition.}
	Two-sided Welch tests were performed for TuD versus MT comparisons at each excitation wavelength. For MT, tests are
	reported for both absorbance branches (bgA and raw).}
	\label{tab:sup_welch_polymer}

	\resizebox{\textwidth}{!}{%
	\begin{tabular}{lllll}
		\\
		\hline
		Excitation & Condition & $p$ value (bgA) & $p$ value (raw) & Interpretation \\
		\hline
		280 & Protein (no Trp)   & $3.018\times 10^{-11}$ & 0.003136 & Significant \\
		280 & Protein + Trp 0.25 & 0.000155463            & 0.335908 & Significant (bgA), Not significant (raw) \\
		280 & Protein + Trp 1.0  & 0.00628070             & 0.534452 & Significant (bgA), Not significant (raw) \\
		295 & Protein (no Trp)   & 0.0459916              & $5.89\times 10^{-7}$ & Significant \\
		295 & Protein + Trp 0.25 & 0.740762               & $2.23\times 10^{-8}$ & Not significant (bgA), Significant (raw) \\
		295 & Protein + Trp 1.0  & 0.269843               & $3.53\times 10^{-10}$ & Not significant (bgA), Significant (raw) \\
		\hline
	\end{tabular}}
\end{table}

\clearpage

\begin{table} 
	\centering
	\caption{\textbf{Theory summary statistics and Welch tests ($n=20$).}
	Quantum yields and percent changes are reported as mean $\pm$ SE across realizations. ``Baseline'' refers to the
	unmodified system. The reported $p$ values are two-sided Welch tests across realizations.}
	\label{tab:sup_theory}

	\small
	\setlength{\tabcolsep}{6pt}
	\renewcommand{\arraystretch}{1.15}

	\begin{tabular}{lcccccc}
		\hline
		\multicolumn{7}{c}{Summary statistics} \\
		\hline
		Condition &
		QY TuD & SE TuD &
		QY MT & SE MT &
		\% change (MT vs TuD) & SE \\
		\hline
		baseline             & 0.135407 & 0.001316 & 0.142052 & 0.0001035 & 4.907\% & 1.022\% \\
		Trp 0.25  & 0.132206 & 0.001009 & 0.138877 & 0.0000778 & 5.046\% & 0.804\% \\
		Trp 1.00             & 0.122602 & 0.000839 & 0.129350 & 0.0000185 & 5.504\% & 0.722\% \\
		\hline
		\multicolumn{7}{c}{Welch tests ($p$ values)} \\
		\hline
		Condition &
		\multicolumn{2}{c}{$p$ (TuD vs MT)} &
		\multicolumn{2}{c}{$p$ (TuD vs baseline)} &
		\multicolumn{2}{c}{$p$ (MT vs baseline)} \\
		\hline
		baseline             & \multicolumn{2}{c}{$7.11\times 10^{-5}$} & \multicolumn{2}{c}{---} & \multicolumn{2}{c}{---} \\
		Trp 0.25  & \multicolumn{2}{c}{$3.69\times 10^{-7}$} & \multicolumn{2}{c}{0.0616} & \multicolumn{2}{c}{$1.38\times 10^{-18}$} \\
		Trp 1.00             & \multicolumn{2}{c}{$1.53\times 10^{-7}$} & \multicolumn{2}{c}{$2.13\times 10^{-9}$} & \multicolumn{2}{c}{$2.23\times 10^{-30}$} \\
		\hline
	\end{tabular}
\end{table}

\clearpage

\paragraph{Caption for Data S1.}
\textbf{Experimental data tables (two CSV files).}
Data S1 contains two combined plate-reader CSV exports, one for the microtubule dataset and one for the tubulin dataset, including the absorbance and fluorescence measurements used for the reported analyses.

\paragraph{Caption for Code S1.}
\textbf{Python analysis script.}
Code S1 contains the Python script used for absorbance tail fitting, quantum-yield calculation, uncertainty propagation, and Welch tests, and generates the summary tables reported in the main text and in Tables~S1--S10.

\end{document}